\documentclass[12pt]{article}
\usepackage{graphicx,multicol}
\usepackage{epstopdf}
\usepackage{amsmath,amssymb,amsfonts,thumbpdf}
\usepackage{amsthm}
\usepackage{enumerate}
\usepackage{bm}
\usepackage{multirow}
\usepackage[authoryear]{natbib}
\usepackage[colorlinks,citecolor=blue]{hyperref}
\usepackage{color}
\usepackage{algorithm}
\usepackage{algorithmic}
\usepackage{float,subcaption}
\usepackage{url, xcolor}
\usepackage{rotating}

\usepackage{natbib}

{\bf}{\it}
\newtheorem{theorem}{Theorem}{\bf}{\it}
{\bf}{\it}

\theoremstyle{definition}
{\bf}{\it}
{\bf}{\it}
{\bf}{\it}
{\bf}{\it}

\newcommand{\blind}{0}

\def\Pr{\mathrm{Pr}}

\def\HLIS{\mathrm{HLIS}}
\def\E{\mathrm{E}}

\def\mFDR{\mathrm{mFDR}}
\def\mFNR{\mathrm{mFNR}}

\usepackage[left=1in, right=1in]{geometry}
\graphicspath{{sim_plot/}}
\allowdisplaybreaks

\begin{document}

\def\spacingset#1{\renewcommand{\baselinestretch}%
{#1}\small\normalsize} \spacingset{1}



  \title{\LARGE\bf Multiple Testing in Genome-Wide Association Studies via Hierarchical Hidden Markov Models}
  \author{Pengfei Wang\footnote{
      Author for Correspondence: Pengfei Wang, E-mail: wangpf0429@dufe.edu.cn. School of Statistics, Dongbei University of Finance and Economics, Dalian 116025, China.} and Zhaofeng Tian
}

  \maketitle

\bigskip
\begin{abstract}
The problems of large-scale multiple testing are often encountered in modern scientific researches. Conventional multiple testing procedures usually suffer considerable loss of testing efficiency due to the lack of consideration of correlations among tests. In fact, the appropriate use of correlation information not only enhances the efficacy of multiple testing but also improves the interpretability of the results. Since the disease- or trait-related single nucleotide polymorphisms (SNPs) often tend to be clustered and exhibit serial correlations, the hidden Markov model (HMM) based multiple testing procedure has been successfully applied in genome-wide association studies (GWAS). It is important to note that modeling the entire chromosome using one HMM is somewhat rough. To overcome this issue, this paper employs the hierarchical hidden Markov model (HHMM) to describe local correlations among tests and develops a multiple testing procedure that can not only automatically divide different class of chromosome regions, but also takes into account local correlations among tests. Theoretically, it is shown that the proposed multiple testing procedure is valid and optimal in some sense. Then a data-driven procedure is developed to mimic the oracle version. Extensive simulations and the real data analysis show that the novel multiple testing procedure outperforms its competitors.
\end{abstract}

\noindent%
\vfill

\newpage
\spacingset{1.45} 
\section{Introduction} \label{sec:intro}
\par
The problems of large scale multiple testing are often encountered in modern scientific researches. For example, in DNA microarray experiments, one needs to test tens of thousands of hypotheses simultaneously to identify genes associated with a certain disease \citep{2001Gene}. In general, large-scale multiple testing poses many statistical challenges compared with a single hypothesis test. For a single test, it is desirable to control the Type I error rate at the pre-specified level $\alpha$ and to minimize the Type II error rate among all $\alpha$-level testing procedures. However, in large-scale multiple testing, controlling the Type I error rate may lead to overly conservative testing procedures. Another challenge is that $p$-values arising from large-scale multiple testing problems often exhibit correlations. For example, in genome-wide association studies (GWAS), since the adjacent genomic loci tend to co-segregate in meiosis, $p$-values generated from GWAS are usually locally dependent. In fact, the proper use of correlation information not only improves the efficiency of multiple testing but also enhances the interpretability of results.

\par
In many practical applications, it is cost-effective to tolerate some false discoveries when the number is only a small proportion of the total number of discoveries. Considering this point, the concept of the false discovery rate (FDR), defined as the expectation of the proportion of false discoveries among all discoveries, was proposed by \cite{benjamini1995controlling}. In fact, the FDR reflects a trade-off between false discoveries and true discoveries. It has become one of the most commonly used control criteria for multiple testing. To date, a series of FDR controlling procedures have been proposed and successfully applied in many scientific fields \citep{benjamini1995controlling,benjamini2000on,genovese2004a}. It should be noted that these conventional multiple testing procedures focused primarily on the validity of the methods and largely ignored the information contained in the correlations among tests. However, a number of studies have shown that the correlations cannot be ignored in multiple testing \citep{owen2005variance,efron2007correlation,schwartzman2011the}. Recently, a growing number of studies have suggested that information about dependent structures should be incorporated into multiple testing to improve the efficiency \citep{shu2015multiple,sun2015false,liu2016multiple}.

\par
Hidden Markov model (HMM), as an effective tool for modeling local correlations, has been successfully applied to large-scale multiple testing \citep{sun2009large}. \cite{sun2009large} suggested to use the local index of significance (LIS), defined as the posterior probability that the null hypothesis is true given all observations, for multiple testing and have shown that it is the optimal statistic for HMM-based multiple testing. To date, a wide range of extensions to the LIS procedure have been proposed \citep{kuan2012integrating,wang2019replicability,wang2019bayesian,cui2021covariate}, among others. For example, \cite{cui2021covariate} extended the LIS procedure to allow for the covariate-adjustment in large-scale multiple testing under dependence.

\par
Since the disease- or trait-related single nucleotide polymorphisms (SNPs) often tend to be clustered and exhibit serial correlations, it is desirable to employ the HMM to describe these correlations. \cite{wei2009multiple} first applied the LIS procedure to GWAS. They suggested that the whole chromosome is regarded as a HMM and the significance threshold is determined by the LIS values of all chromosomes. It is important to note that modeling the entire chromosome using one HMM is somewhat rough. In view of this, \cite{xiao2013large} suggested that different regions of the chromosome should be modeled separately and developed a procedure, called region-specific pooled LIS (RSPLIS) procedure, which finds multiple change points on chromosomes by using the dynamic programming (DP) algorithm. However, the process of finding change points by the RSPLIS procedure is separate from the multiple testing process, and the change points found are poorly interpreted.

\par
In this article, we propose a multiple testing procedure, called hierarchical local index of significance (HLIS) procedure, that can not only automatically divide different regions, but also characterize the local dependence among tests. Specifically, a hidden layer is added to the HMM, and the state of the added hidden layer variables is used to indicate the type of the divided regions. The identification of different regions is achieved by estimating the state of the newly added layer. To characterize the local dependence among tests, we further assume that the added hidden layer variables follow a block-wise Markov chain. In essence, under the above model assumptions, observations and two-layer hidden variables constitute a hierarchical hidden Markov model (HHMM), which has been successfully applied to the chromatin-state analysis \citep{Marco2017multi}. Based on the HHMM, we develop the HLIS procedure and show that it is valid and optimal, that is, the HLIS procedure is capable of controlling the marginal false discovery rate (mFDR) at the pre-specified level $\alpha$ and minimizes the marginal false non-discovery rate (mFNR) among all $\alpha$-level testing procedures. In practice, the parameters of the HHMM are usually unknown. We utilize expectation-maximization (EM) algorithm to calculate the maximum likelihood estimations (MLEs) of the parameters of the HHMM and employ the nonparametric Gaussian kernel method \citep{Silverman2018density} to estimate the probability density function (pdf) corresponding to the non-null. A data-driven HLIS procedure is developed to mimic the oracle version. Extensive simulations demonstrate that the HLIS procedure is valid and is capable of identifying different types of regions accurately. Both simulations and the real data analysis illustrate that the HLIS procedure provides a powerful tool for multiple testing in GWAS.

\par
The rest of this paper is organized as follows. Section 2 starts with a brief description of the HHMM. Then we presents the oracle HLIS procedure based on the HHMM and its corresponding theoretical results. Subsequently, the data-driven HLIS procedure to mimic the oracle version and the algorithm for implementing the HLIS procedure are given. Section 3 presents exhaustive simulations in various parameter settings. In Section 4, we apply the HLIS procedure to schizophrenia (SCZ) data analysis. Some discussions and suggestions for future work are summarized in Section 5. The proofs of all theorems are placed in the Appendix.

\newpage

\section{Statistical Methods}\label{sec:model}

\subsection{The hierarchical hidden Markov model}\label{sec-2.1}

\par
Suppose that the problem of GWAS we are interested in has $m$ hypotheses to be tested simultaneously. Let $\{\theta_i\}^m_{i=1}$ be the underlying states of hypotheses, where $\theta_i=1$ means that the $i$th null hypothesis is not true, that is, the $i$th SNP is associated with the corresponding disease or trait, and $\theta_i=0$ otherwise. Let $\{\eta_i\}^m_{i=1}$ be the underlying types of the divided regions where $\eta_i=k$ means that the $i$th SNP is on the $k$th region of the chromosome, for $k=1,\cdots,K$. Let $\{z_i\}^m_{i=1}$ be the sequence of observations, where $z_i$ can be the $z$-value or the statistic for the $i$th test. Denote by $Z_i$ the random variable with respect to $z_i$.

\par
To characterize the local dependence among tests, assume that the sequence of underlying types $\{\eta_i\}^m_{i=1}$ follows a block-wise Markov chain with the initial probabilities:
\[
        \Pr(\eta_1=k) = \pi_k, \text{~for~}k=1, \cdots, K, \eqno{(1)}
\]
and the transition probabilities:
$$\Pr(\eta_{i+1}=l \mid \eta_i=k) =
\begin{cases}
\delta_{kl},~& \text{~if~}i\text{~is~not~a~multiple~of~}S,\\
b_{kl},~& \text{~if~}i\text{~is~a~multiple~of~}S,
\end{cases}\eqno{(2)}
$$
where $\delta_{kl}$ is the Kronecker delta and $S$ is the size of the block, for $k,l=1, \cdots, K$. In essence, these transition probabilities make the change of the value of $\eta_i$ only occur at the end of the block of size $S$. Moreover, assume that the null hypothesis state $\theta_i$ depends on the previous state $\theta_{i-1}$ and the current type $\eta_i$ with the transition probabilities:
\[
        \Pr(\theta_i=q \mid \theta_{i-1}=p, \eta_i=k) = a_{pq}(k), \text{~for~}k=1, \cdots, K, \text{~and~} p, q=0, 1, \eqno{(3)}
\]
and assume that the initial probabilities are:
\[
        \Pr(\theta_1=p \mid \eta_1=k) = c_p(k), \text{~for~}k=1, \cdots, K, \text{~and~} p=0, 1.\eqno{(4)}
\]
In the same way as \cite{sun2009large}, we assume that the random variables $\{Z_i\}^m_{i=1}$ are conditionally independent given the underlying states of hypotheses $\{\theta_i\}^m_{i=1}$, that is,
\[
       \Pr(\{Z_i\}^m_{i=1}\mid\{\theta_i\}^m_{i=1})=\prod^m_{i=1} \Pr(Z_i\mid\theta_i). \eqno{(5)}
\]
Using the commonly used two-component mixture model \citep{efron2001empirical}, we have
\[
         Z_i \mid \theta_i \sim  (1-\theta_i)F_0 + \theta_i F_1, \eqno{(6)}
\]
where $F_0$ and $F_1$ denote the cumulative distribution functions (cdfs) given $\theta_i=0$ and $\theta_i=1$, respectively. The pdfs corresponding to $F_0$ and $F_1$ are denoted by $f_0$ and $f_1$, respectively. A schematic diagram of the HHMM for large-scale multiple testing is presented in Figure 1. To simplify the notation, denote by $\boldsymbol{\pi}=(\pi_1, \cdots, \pi_K)$ the initial probabilities, $\boldsymbol{c}=(c_0(1), \cdots, c_0(K); c_1(1), \cdots, c_1(K))$ the initial conditional probabilities, $\boldsymbol{\mathcal{A}}=(\mathcal{A}_1, \cdots, \mathcal{A}_K)$ the transition probability matrices, where $\mathcal{A}_k = \{a_{pq}(k)\}_{2\times2}$ for $k=1, \cdots, K$, $\boldsymbol{\mathcal{B}} = \{b_{kl}\}_{K\times K}$ the block-wise transition probability matrix, $\boldsymbol{\mathcal{F}}=(f_0, f_1)$ the pdfs given $\theta_i=0$ and $\theta_i=1$, and $\boldsymbol{\vartheta}=(\boldsymbol{\pi}, \boldsymbol{c}, \boldsymbol{\mathcal{A}}, \boldsymbol{\mathcal{B}}, \boldsymbol{\mathcal{F}})$ the parameters of the HHMM.

\begin{figure}[H]
\centering
\includegraphics[height=2.2in,width=5.6in]{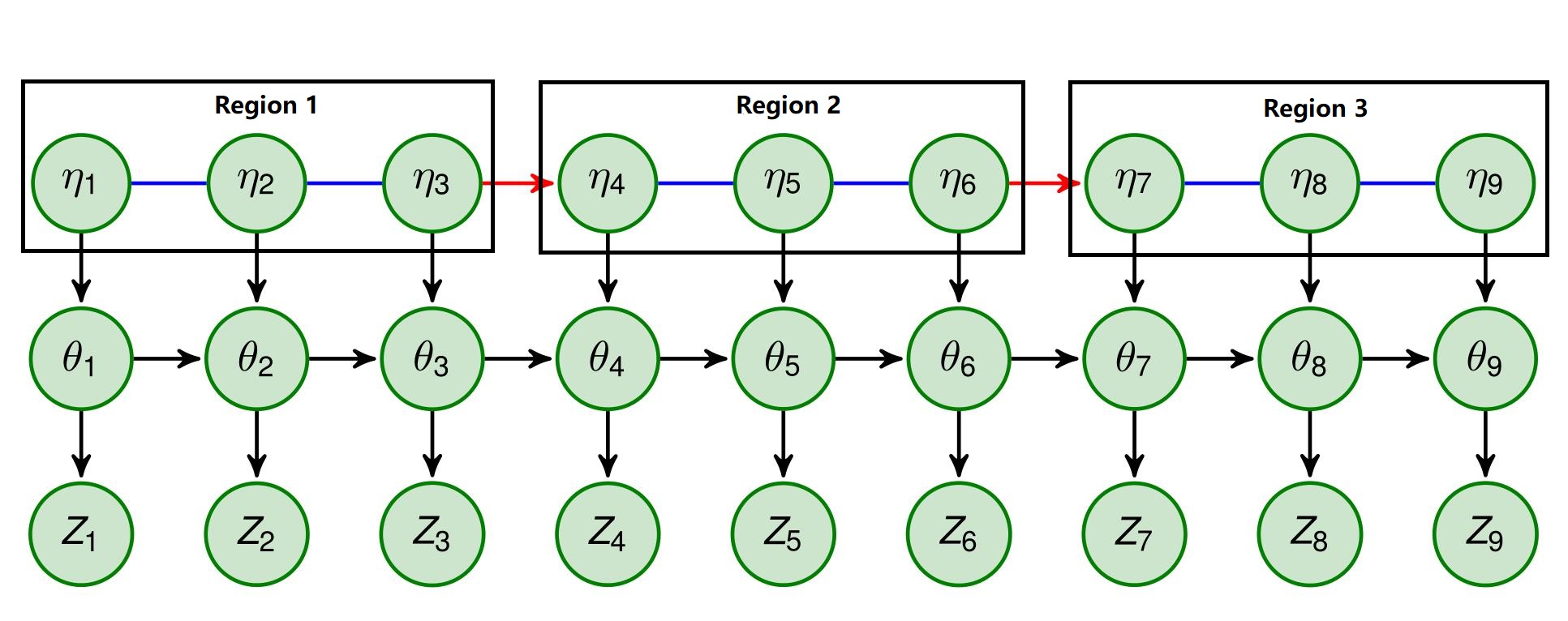}
\caption{\footnotesize{A schematic diagram of the HHMM}}
\end{figure}

\subsection{The HLIS procedure}

\par
In this section, we first consider the case where the parameters of the HHMM are known. Then we propose the oracle HLIS procedure and provide some theoretical results. The forward-backward algorithm for calculating the HLIS statistic is displayed in Subsection 2.2.2. When the parameters of the HHMM are unknown, we introduce the data-driven HLIS procedure and then provide the EM algorithm for fitting the HHMM.

\subsubsection{The oracle HLIS procedure}\label{sec-2.2}

\par
Based on the HHMM, we define hierarchical local index of significance, HLIS, for the $j$th test as:
\[
       \HLIS_j = \Pr(\theta_j=0 \mid \{z_{i}\}^m_{i=1}, \boldsymbol{\vartheta}).
\]

Although the definitions of HLIS and LIS look the same in form, they have many significant differences in essence.
\begin{enumerate}[~~~~~~(a)]
\item The definitions of HLIS and LIS are based on different local dependency models.
\item HLIS can be expanded into $K$ parts, namely,
\[
        \HLIS_j = \sum^K_{k=1} \Pr(\theta_j=0 \mid \eta_j=k, \{z_{i}\}^m_{i=1}, \boldsymbol{\vartheta})\Pr(\eta_j=k \mid \{z_{i}\}^m_{i=1}, \boldsymbol{\vartheta}),
\]
where $\Pr(\theta_j=0 \mid \eta_j=k, \{z_{i}\}^m_{i=1}, \boldsymbol{\vartheta})$ is the $j$-th LIS statistic for the chromosome region of Type $k$. Thus the HLIS statistic can be viewed as a weighted LIS statistic, where the weights are the posterior probabilities of the region categories given all observations. In particular, if there is only one type of region, that is, $K=1$, then the HLIS statistic will degenerate to the LIS statistic.
\item The HLIS statistic contains information on the type of the chromosome region, whereas the LIS statistic does not.
\end{enumerate}

\par
For ease of exposition, $\{Z_i\}^m_{i=1}$ is abbreviated as $\boldsymbol{Z}$. Then we denote $\Pr(\theta_j=0 \mid \{Z_i\}^m_{i=1}, \boldsymbol{\vartheta})$ by $\HLIS_j\left(\boldsymbol{Z}\right)$. Let ${\bm\delta}(\HLIS(\boldsymbol{Z}),c)=\left\{I(\HLIS_j\left(\boldsymbol{Z}\right)< c)\right\}^m_{j=1}$ be the decision rule with respect to the HLIS statistic, where $I(\cdot)$ is an indicator function and $c$ is a cut-off. In such a case, $\HLIS_j\left(\boldsymbol{Z}\right)< c$ implies that the $j$th null hypothesis is rejected by the decision rule ${\bm\delta}(\HLIS(\boldsymbol{Z}),c)$, and $\HLIS_j\left(\boldsymbol{Z}\right) \geq c$ otherwise. Next, we start by showing that there exists a cut-off $c_{\alpha}$ such that
\[
        \mFDR({\bm\delta}(\HLIS(\boldsymbol{Z}), c_{\alpha}))=\alpha,
\]
where $\mFDR({\bm\delta}(T, c))$ is the mFDR corresponding to ${\bm\delta}(T, c)$. Then we will show that the decision rule ${\bm\delta}(\HLIS(\boldsymbol{Z}), c_{\alpha})$ is optimal, that is, ${\bm\delta}(\HLIS(\boldsymbol{Z}), c_{\alpha})$ yields the smallest mFNR among all decision rules with the mFDR controlled at $\alpha$. In practice, the cut-off $c_{\alpha}$ is usually unknown. We further give a strategy to select an appropriate cut-off $c^{*}$ so that the FDR level of the decision rule ${\bm\delta}(\HLIS(\boldsymbol{Z}), c^{*})$ is controlled at $\alpha$. The novel decision rule ${\bm\delta}(\HLIS(\boldsymbol{Z}), c^{*})$ is referred to as the HLIS procedure. Note that $\mFDR({\bm\delta}(T, c))=\mathrm{FDR}({\bm\delta}(T, c))+O(1/\sqrt{m})$ under some mild conditions \citep{genovese2002operating}. It follows that the HLIS procedure is asymptotically optimal in the sense of controlling the mFDR and minimizing the mFNR. The following theorem (Theorem 1) shows the existence of such a $c_{\alpha}$.

\begin{theorem}
Consider the hierarchical hidden Markov model (1)-(6). Let ${\bm\delta}(\HLIS(\boldsymbol{Z}),c)$ be the decision rule corresponding to the HLIS statistic, then there exists a cut-off $c_{\alpha}$ such that
$$\mFDR\left({\bm\delta}(\HLIS(\boldsymbol{Z}), c_{\alpha})\right)=\alpha,$$
that is,
\[
        \frac{\mathrm{E}\left[\sum\limits^m_{j=1}I(\HLIS_j(\boldsymbol{Z})<c_{\alpha})(1-\theta_{j})\right]}{\mathrm{E}\left[\sum\limits^m_{j=1}I(\HLIS_j(\boldsymbol{Z})<c_{\alpha})\right]}=\alpha.
\]
\end{theorem}

\par
The next theorem (Theorem 2) illustrates that the decision rule ${\bm\delta}(\HLIS(\boldsymbol{Z}), c_{\alpha})$ is optimal in the sense of controlling the mFDR and minimizing the mFNR.

\begin{theorem}
Consider the hierarchical hidden Markov model (1)-(6). Assume that the decision rule ${\bm\delta}(\HLIS(\boldsymbol{Z}),c_{\alpha})$ satisfies the condition $\mFDR\left({\bm\delta}(\HLIS(\boldsymbol{Z}), c_{\alpha})\right)=\alpha$, then it yields the smallest mFNR among all decision rules with the mFDR controlled at $\alpha$. That is, for any decision rule ${\bm\delta}(T(\boldsymbol{Z}),c)=\left\{I(T_j\left(\boldsymbol{Z}\right)< c)\right\}^m_{i=1}$ with $\mFDR({\bm\delta}(T(\boldsymbol{Z}),c))\leq\alpha$, we have
\[
        \frac{\mathrm{E}\left\{\sum\limits^m_{j=1}\left[1-I(\HLIS_j(\boldsymbol{Z})<c_{\alpha})\right]\theta_{j}\right\}}{\mathrm{E}\left\{\sum\limits^m_{j=1}\left[1-I(\HLIS_j(\boldsymbol{Z})<c_{\alpha})\right]\right\}}\leq \frac{\mathrm{E}\left\{\sum\limits^m_{j=1}\left[1-I(T_j(\boldsymbol{Z})<c)\right]\theta_{j}\right\}}{\mathrm{E}\left\{\sum\limits^m_{j=1}\left[1-I(T_j(\boldsymbol{Z})<c)\right]\right\}}.
\]
\end{theorem}

\par
In practice, however, $c_{\alpha}$ is usually unknown and needs to be estimated. Following the general idea used in \cite{genovese2004a}, \cite{Newton2004Detecting} and \cite{sun2007oracle}, we provide a strategy to select an appropriate cut-off $c^{*}$ so that the FDR level of the decision rule ${\bm\delta}(\HLIS(\boldsymbol{Z}), c^{*})$ is controlled at $\alpha$. Specifically, denote by $\HLIS_{(1)},\HLIS_{(2)},...,\HLIS_{(m)}$ the ordered HLIS statistics and $H_{(1)},H_{(2)},...,H_{(m)}$ the corresponding null hypotheses. Then the oracle HLIS procedure operates as follows
\[
        \text{Let}~l=\max\left\{i:\frac{1}{i}\sum\limits^i_{j=1}\HLIS_{(j)}\leq\alpha\right\};~\text{then~reject~all~} H_{(j)}, \text{~for~} j=1,...,l.\eqno{(7)}
\]
In such a case, $c^{*}$ can take any value in the interval $\left(\HLIS_{(l)}, \HLIS_{(l+1)}\right]$. On the other hand, the oracle HLIS procedure is the decision rule ${\bm\delta}(\HLIS(\boldsymbol{Z}), c^{*})$ with $\HLIS_{(l)}<c^{*}\leq\HLIS_{(l+1)}$. The next theorem (Theorem 3) states that the oracle HLIS procedure is capable of controlling the FDR at level $\alpha$.

\begin{theorem}
Consider the hierarchical hidden Markov model (1)-(6). The oracle HLIS procedure (7) controls the FDR at level $\alpha$.
\end{theorem}

\subsubsection{The algorithm for calculating the HLIS statistic}

By using the forward-backward algorithm \citep{baum1970a} with minor modifications, the HLIS statistic can be calculated efficiently. Specifically, the HLIS statistic for the $j$-th test can be expressed as:
\[
   \HLIS_j=\frac{\sum^K_{k=1}\alpha_j(0, k)\beta_j(0, k)}{\sum^1_{p=0}\sum^K_{k=1}\alpha_j(p, k)\beta_j(p, k)},
\]
where $\alpha_j(p, k)=\Pr(\theta_j=p, \eta_j=k, \{z_{i}\}^j_{i=1} \mid \boldsymbol{\vartheta})$ and $\beta_j(p, k)=\Pr(\{z_{i}\}^m_{i=j+1}\mid \theta_j=p, \eta_j=k, \boldsymbol{\vartheta})$, for $p=0, 1$ and $k=1, \cdots, K$, are the forward variable and the backward variable, respectively. By some mathematical derivations, we have
\[
   \alpha_{j+1}(p, k) = f_p(z_{j+1})\sum^1_{q=0}\sum^K_{l=1} \left[\alpha_j(q, l) a_{qp}(k)\delta_{lk}^{s(j)}b_{lk}^{1-s(j)}\right],
\]
and
\[
   \beta_j(p, k) = \sum^1_{q=0}\sum^K_{l=1} \left[f_{q} (z_{j+1}) \beta_{j+1}(q, l) a_{pq}(l)\delta_{kl}^{s(j)}b_{kl}^{1-s(j)} \right],
\]
where $\alpha_1(p, k) = \pi_k c_p(k) f_p(z_1)$, $\beta_m(p, k)=1$, for $p=0,1$ and $k=1, \cdots, K$, and
$$s(j) =
\begin{cases}
1,~& \text{~if~}j\text{~is~not~a~multiple~of~}S,\\
0,~& \text{~if~}j\text{~is~a~multiple~of~}S.
\end{cases}
$$

\subsubsection{The data-driven HLIS Procedure}

In practice, the parameters of the HHMM are usually unknown. We employ the EM algorithm to fit the HHMM. By replacing the parameters of the HHMM with their maximum likelihood estimations, we can obtain the plug-in $\widehat{\HLIS}_j, j=1,...,m$. Denote by $\widehat{\HLIS}_{(1)},\widehat{\HLIS}_{(2)},...,\widehat{\HLIS}_{(m)}$ the ordered plug-in HLIS statistics and $H_{(1)},H_{(2)},...,H_{(m)}$ the corresponding null hypotheses. Then the data-driven HLIS procedure operates as follows
\[
        \text{Let}~l=\max\left\{i:\frac{1}{i}\sum\limits^i_{j=1}\widehat{\HLIS}_{(j)}\leq\alpha\right\};~\text{then~reject~all~} H_{(j)}, \text{~for~} j=1, \cdots, l.\eqno{(8)}
\]
Next, we provide the detailed EM algorithm for model fitting.

\subsubsection{The algorithm for fitting the HHMM}

\par
To fit the HHMM, the parameters are estimated by using the expectation-maximization (EM) algorithm. It should be noted that the non-null pdf $f_1$ is unknown in practice. \cite{sun2009large} suggested to use the normal mixture model to fit the alternatives and to employ Bayesian information criterion (BIC) to choose the number of mixture components $L$. In many circumstances, however, the pdf $f_1$ may be too complex to use the mixed normal approximation and the method for choosing $L$ may be computationally intensive. To overcome these limitations, we utilize the nonparametric Gaussian kernel density estimation (Silverman, 2018) to estimate the non-null pdf $f_1$. Specifically, $f_1$ can be estimated by
\[
        \widehat{f_1}(z) = \dfrac{\sum^m_{j=1}\gamma_j(1)K_h(z-z_j)}{\sum^m_{j=1}\gamma_j(1)},
\]
where $\gamma_j(1) = \Pr(\theta_j=1 \mid \{z_i\}^m_{i=1}, \boldsymbol{\vartheta})$, $K_h(\cdot)$ is the Gaussian kernel, and $h$ is the bandwidth.

\begin{algorithm}[htp]
  \begin{itemize} \setlength\itemsep{-0.5em}
  \item[ ] \textbf{Input:} the observations $\{z_i\}^m_{i=1}$.
  \item[ ] \textbf{Output:} the parameters $\boldsymbol{\vartheta}=(\boldsymbol{\pi}, \boldsymbol{c}, \boldsymbol{\mathcal{A}}, \boldsymbol{\mathcal{B}}, \boldsymbol{\mathcal{F}})$ of the HHMM.
  \item[ ]  {S{\footnotesize TEP} 1. Initialize}
  $\boldsymbol{\vartheta}^{(0)}=(\boldsymbol{\pi}^{(0)}, \boldsymbol{c}^{(0)}, \boldsymbol{\mathcal{A}}^{(0)}, \boldsymbol{\mathcal{B}}^{(0)}, \boldsymbol{\mathcal{F}}^{(0)})$
  \item[ ]  {S{\footnotesize TEP} 2 (E-Step). Calculate the following variables:}\\[-1cm]
    \begin{enumerate}[(a)]
       \item $\alpha^{(t-1)}_j(p, k)=\Pr(\theta_j=p, \eta_j=k, \{z_{i}\}^j_{i=1} \mid \boldsymbol{\vartheta}^{(t-1)})$;
       \item $\beta^{(t-1)}_j(p, k)=\Pr(\{z_{i}\}^m_{i=j+1}\mid \theta_j=p, \eta_j=k, \boldsymbol{\vartheta}^{(t-1)})$;
       \item $\xi^{(t-1)}_j(p, q, k, l)=\Pr(\theta_j=p, \theta_{j+1}=q, \eta_j=k, \eta_{j+1}=l \mid \{z_i\}^m_{i=1}, \boldsymbol{\vartheta}^{(t-1)})$;
       \item $\phi^{(t-1)}_1(k)=\Pr(\eta_1=k \mid \{z_i\}^m_{i=1}, \boldsymbol{\vartheta}^{(t-1)})$;
       \item $\nu^{(t-1)}_j(k, l)=\Pr(\eta_j=k, \eta_{j+1}=l \mid \{z_i\}^m_{i=1}, \boldsymbol{\vartheta}^{(t-1)})$;
       \item $\zeta^{(t-1)}_j(p, q, k)=\Pr(\theta_j=p, \theta_{j+1}=q, \eta_{j+1}=k \mid \{z_i\}^m_{i=1}, \boldsymbol{\vartheta}^{(t-1)})$;
       \item $\rho^{(t-1)}_1(p, k)=\Pr(\theta_1=p, \eta_1=k \mid \{z_i\}^m_{i=1}, \boldsymbol{\vartheta}^{(t-1)})$;
       \item $\gamma^{(t-1)}_j(p)=\Pr(\theta_j=p \mid \{z_i\}^m_{i=1}, \boldsymbol{\vartheta}^{(t-1)})$,
     \end{enumerate}\vskip -0.3cm
      \qquad for $p, q=0, 1$, $k, l=1, \cdots, K$, and $j=1, \cdots, m$.
  \item[ ]  {S{\footnotesize TEP} 3 (M-Step). Update the following parameters:}\\[-1cm]
    \begin{enumerate}[(a)]
       \item $\pi^{(t)}_k = \phi^{(t-1)}_1(k)$;
       \item $c^{(t)}_p(k) = \rho^{(t-1)}_1(p, k)/\sum^1_{q=0}\rho^{(t-1)}_1(q, k)$;
       \item $a^{(t)}_{pq}(k) = \sum^{m-1}_{j=1}\zeta^{(t-1)}_j(p, q, k)/\sum^{m-1}_{j=1}\sum^1_{r=0}\zeta^{(t-1)}_j(p, r, k)$;
       \item $b^{(t)}_{kl} = \sum_{\{j: j\mid S = 0\}}\xi^{(t-1)}_j(k, l)/\sum_{\{j: j\mid S = 0\}}\sum^K_{s=1}\xi^{(t-1)}_j(k, s)$;
       \item $f^{(t)}_p(z) = \sum^m_{j=1}\gamma^{(t-1)}_j(p)K_h(z-z_j)/\sum^m_{j=1}\gamma^{(t-1)}_j(p)$,
    \end{enumerate}
       \qquad for $p, q=0, 1$, and $k, l=1, \cdots, K$.
  \item[ ]  {S{\footnotesize TEP} 4. Iterate the S{\footnotesize TEP} 2 and S{\footnotesize TEP} 3 in turn.}
  \end{itemize}
  \caption{{\bf EM} algorithm for estimating parameters $\boldsymbol{\vartheta}$}
  \label{alg:1}
\end{algorithm}

\section{Simulation Studies}
\label{sec-3}

To evaluate the numerical performance of the HLIS procedure, we conduct extensive simulation studies. The simulations are divided into two parts according to the different mechanisms of generating simulated data. In the first part of simulations, the simulated data is generated from the HHMM under various parameter settings, while the second part of the simulated data is generated from a more realistic data. We compare the HLIS procedure against three state-of-the-art procedures for multiple testing: (1) the BH procedure \citep{benjamini1995controlling}; (2) the Lfdr procedure \citep{efron2002empirical}; and (3) the LIS procedure \citep{sun2009large}. The R code for implementing the HLIS procedure is available from https://github.com/wpf19890429/Multiple-Testing-in-Genome-Wide-Association-Studies-via-Hierarchical-Hidden-Markov-Models.

\subsection{Simulation I}\label{sec-3.1}

\par
In Simulation I, the simulated data are generated from the HHMM described in Subsection 2.1. According to the number of chromosome region types, Simulation I is divided into two cases: $K=2$ and $K=3$. The observations $\{z_i\}^m_{i=1}$ are generated from the two-component mixture model (6), where $F_0\sim N(0, 1)$ and $F_1\sim \lambda N(\mu_1, 1)+(1-\lambda) N(2, 1)$. Without loss of generality, $m$ and $S$ are fixed at $9000$ and $30$, respectively. All simulation results are based on $100$ repetitions.

\par
{\bf Case 1 ($K=2$):}

\par
In Case 1, the underlying states of chromosome region types $\{\eta_i\}^m_{i=1}$ are generated from a block-wise Markov chain with the initial probabilities: $\boldsymbol{\pi}=(0.5, 0.5)$, and the block-wise transition probability matrix:
\begin{equation*}       
\boldsymbol{\mathcal{B}} =
\left(                 
  \begin{array}{cc}   
    0.9    & 0.1 \\  
    b_{21} & 1-b_{21} \\  
  \end{array}
\right).                 
\end{equation*}
The underlying states of null hypotheses $\{\theta_i\}^m_{i=1}$ are generated from a process with the initial probabilities:
\begin{equation*}       
\boldsymbol{c} =
\left(                 
  \begin{array}{cc}   
    0.5 & 0.5 \\  
    0.5 & 0.5 \\  
  \end{array}
\right),               
\end{equation*}
and the transition probability matrices:
\[
    \mathcal{A}_1 = \left( {\begin{array}{*{20}c} 0.9 & 0.1 \\ a_{10}(1) & 1-a_{10}(1) \\\end{array}} \right), \quad \mathcal{A}_2 = \left( {\begin{array}{*{20}c} 0.3 & 0.7 \\ 0.7 & 0.3 \\\end{array}} \right).
\]
We conduct a series of simulations under the following parameter settings.

\par
{\bf Setting 1:} fix $\lambda=1$, $\mu_1=2$, $b_{21}=0.1$ and change $a_{10}(1)$ from $0.1$ to $0.2$.

\par
{\bf Setting 2:} fix $\lambda=1$, $\mu_1=2$, $a_{10}(1)=0.2$ and change $b_{21}$ from $0.1$ to $0.2$.

\par
{\bf Setting 3:} fix $\lambda=1$, $b_{21}=0.1$, $a_{10}(1)=0.2$ and change $\mu_1$ from $1$ to $2$.

\par
{\bf Setting 4:} fix $\lambda=0.5$, $\mu_1=1$, $b_{21}=0.1$ and change $a_{10}(1)$ from $0.1$ to $0.2$.

\par
{\bf Setting 5:} fix $\lambda=0.5$, $\mu_1=1.5$, $a_{10}(1)=0.2$ and change $b_{21}$ from $0.1$ to $0.2$.

\par
{\bf Setting 6:} fix $\lambda=0.5$, $b_{21}=0.1$, $a_{10}(1)=0.2$ and change $\mu_1$ from $1$ to $2$.

\par
The detailed simulation results for Settings 1-3 and 4-6 are presented in Figures 2 and 3, respectively. From Panels (a), (c) and (e) of Figure 2, we can observe that: (1) the oracle HLIS procedure, the data-driven HLIS procedure and the Lfdr procedure can control the FDR well in the neighborhood of $0.1$; (2) the BH procedure is somewhat conservative; (3) the LIS procedure is not valid in Settings 1-3. From Panels (b), (d) and (f) of Figure 2, we can find that: (1) the FNR values yielded by the oracle HLIS procedure and the data-driven HLIS procedure are quite close, which indicates that the data-driven HLIS procedure can mimic the oracle version quite well; (2) both HLIS procedures have the smallest FNR, followed by the LIS procedure, the Lfdr procedure and the BH procedure; (3) for $\mu_1=1$ in Setting 3, the FNR values of all procedures are relatively close, which is due to the weak signal that makes all procedures less effective. Note that the larger the value of $\mu_1$, the stronger the signal, so it is straightforward to understand that the FNR decreases as $\mu_1$ increases. From Figure 3, we can obtain similar conclusions, which are not repeated here. It is important to highlight that the LIS procedure has the the smallest FNR in Settings 4-6, which is due in part to its invalid FDR control.

\begin{figure}[htp]
  \centering
  \begin{subfigure}{0.48\textwidth}
    \includegraphics[width=\textwidth,height=75mm]{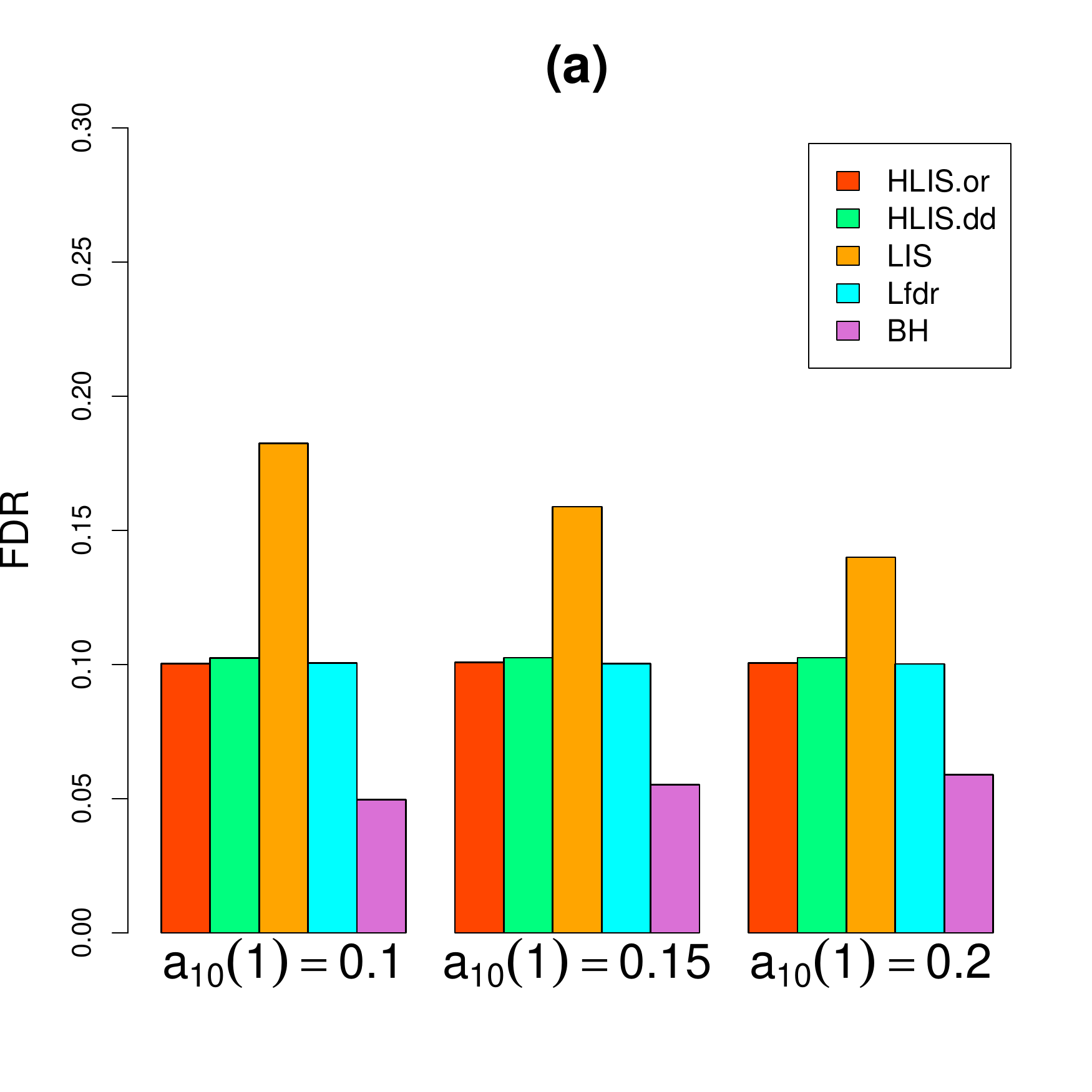}
  \end{subfigure}
  \begin{subfigure}{0.48\textwidth}
    \includegraphics[width=\textwidth,height=75mm]{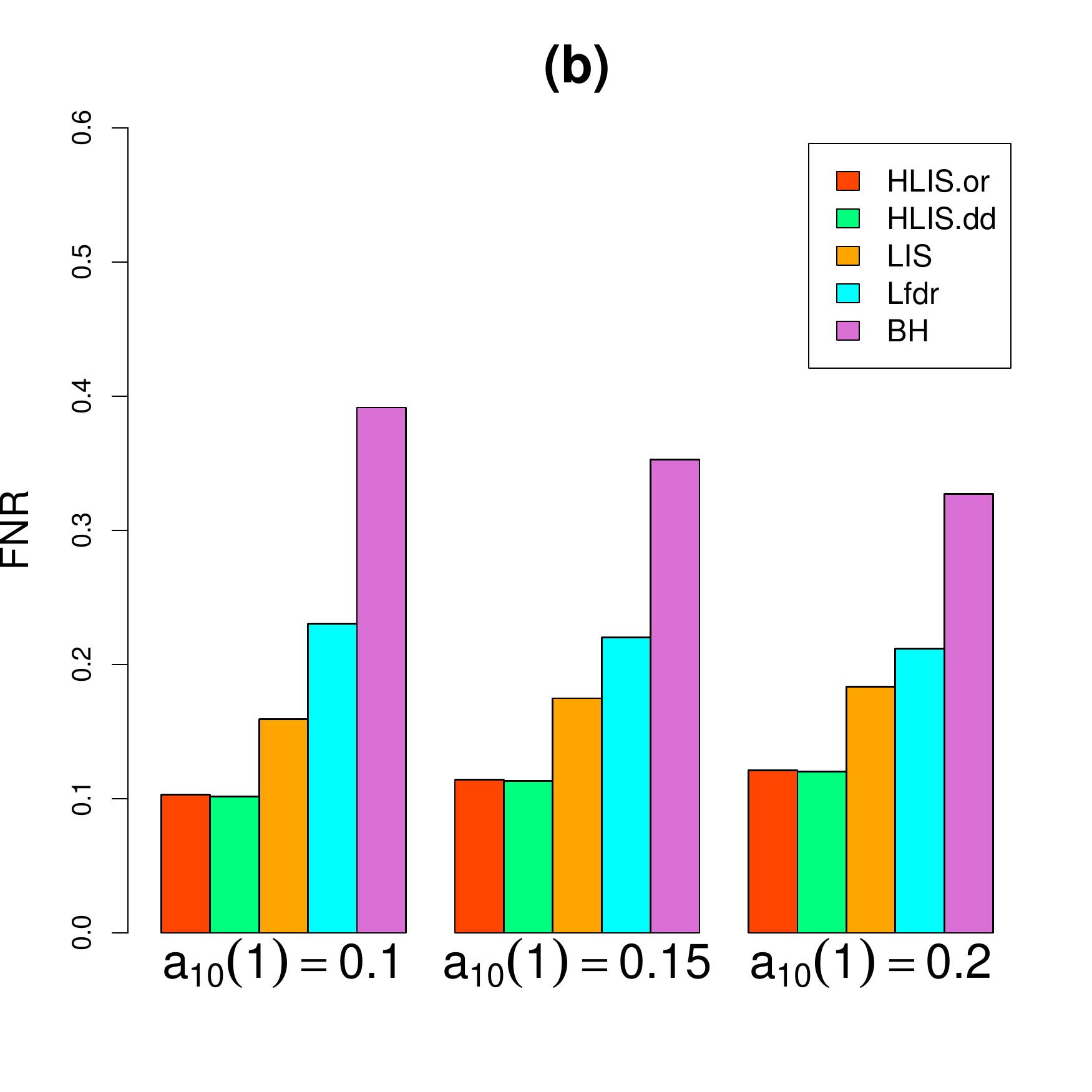}
  \end{subfigure}\\[-7mm]
  \begin{subfigure}{0.48\textwidth}
    \includegraphics[width=\textwidth,height=75mm]{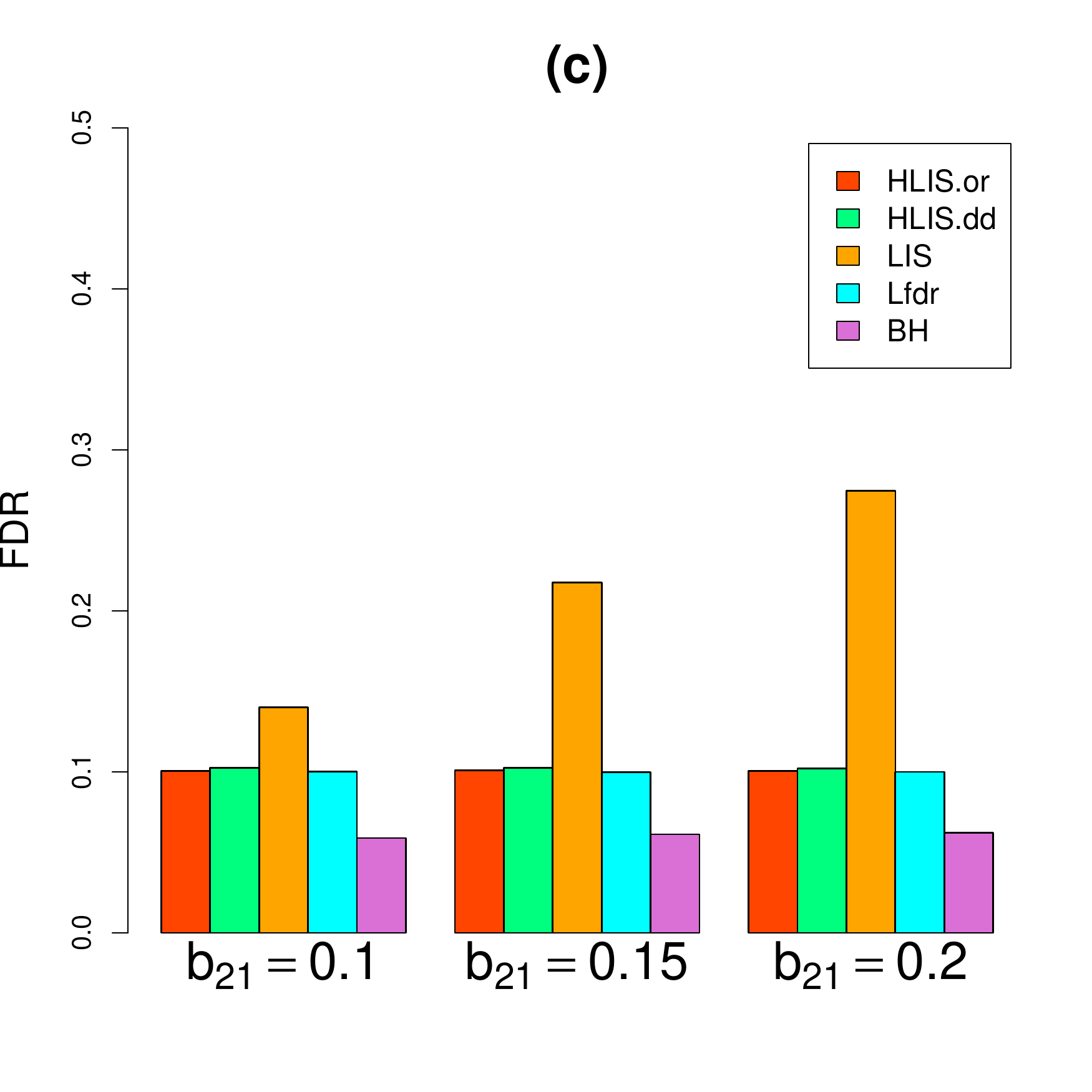}
  \end{subfigure}
  \begin{subfigure}{0.48\textwidth}
    \includegraphics[width=\textwidth,height=75mm]{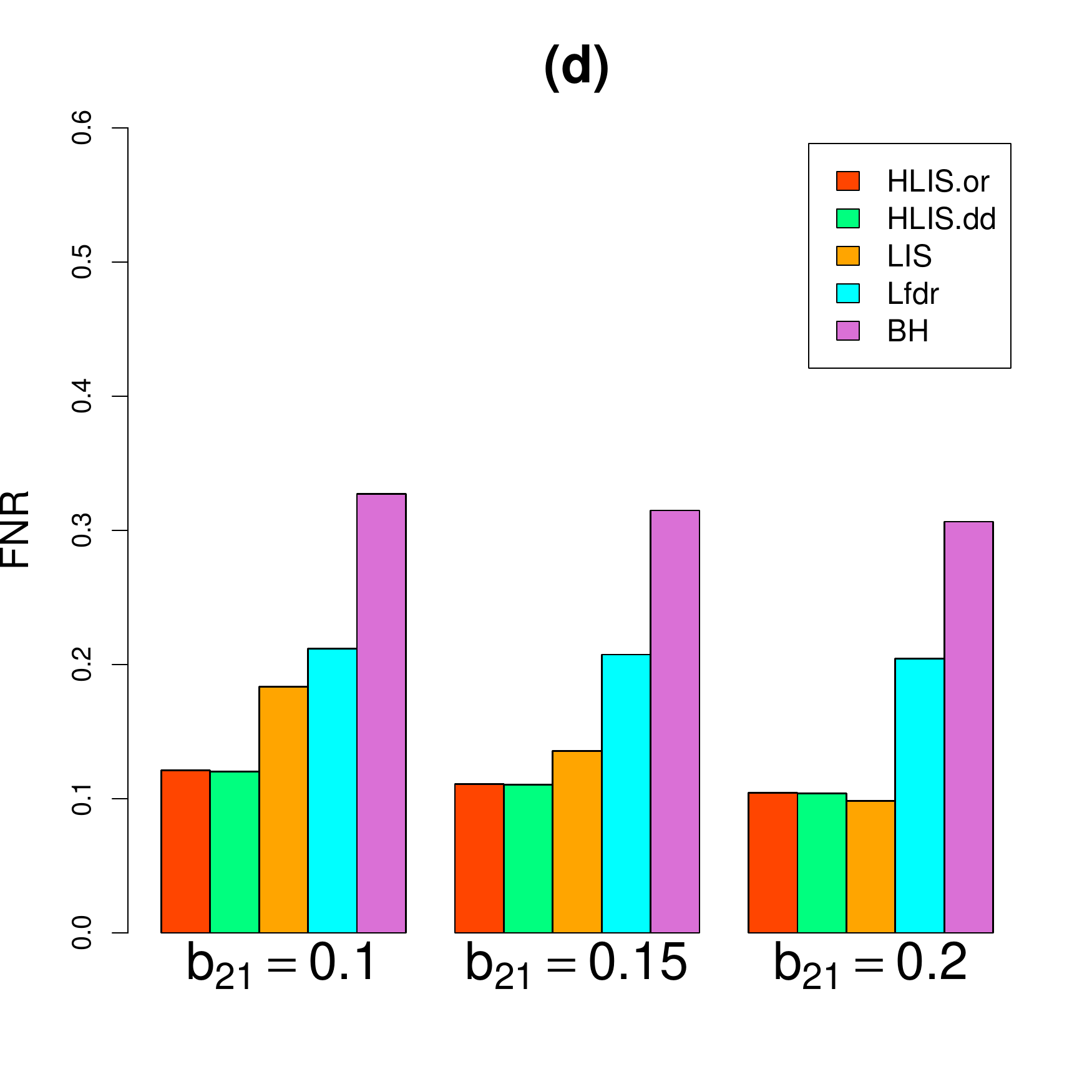}
  \end{subfigure}\\[-7mm]
  \begin{subfigure}{0.48\textwidth}
    \includegraphics[width=\textwidth,height=75mm]{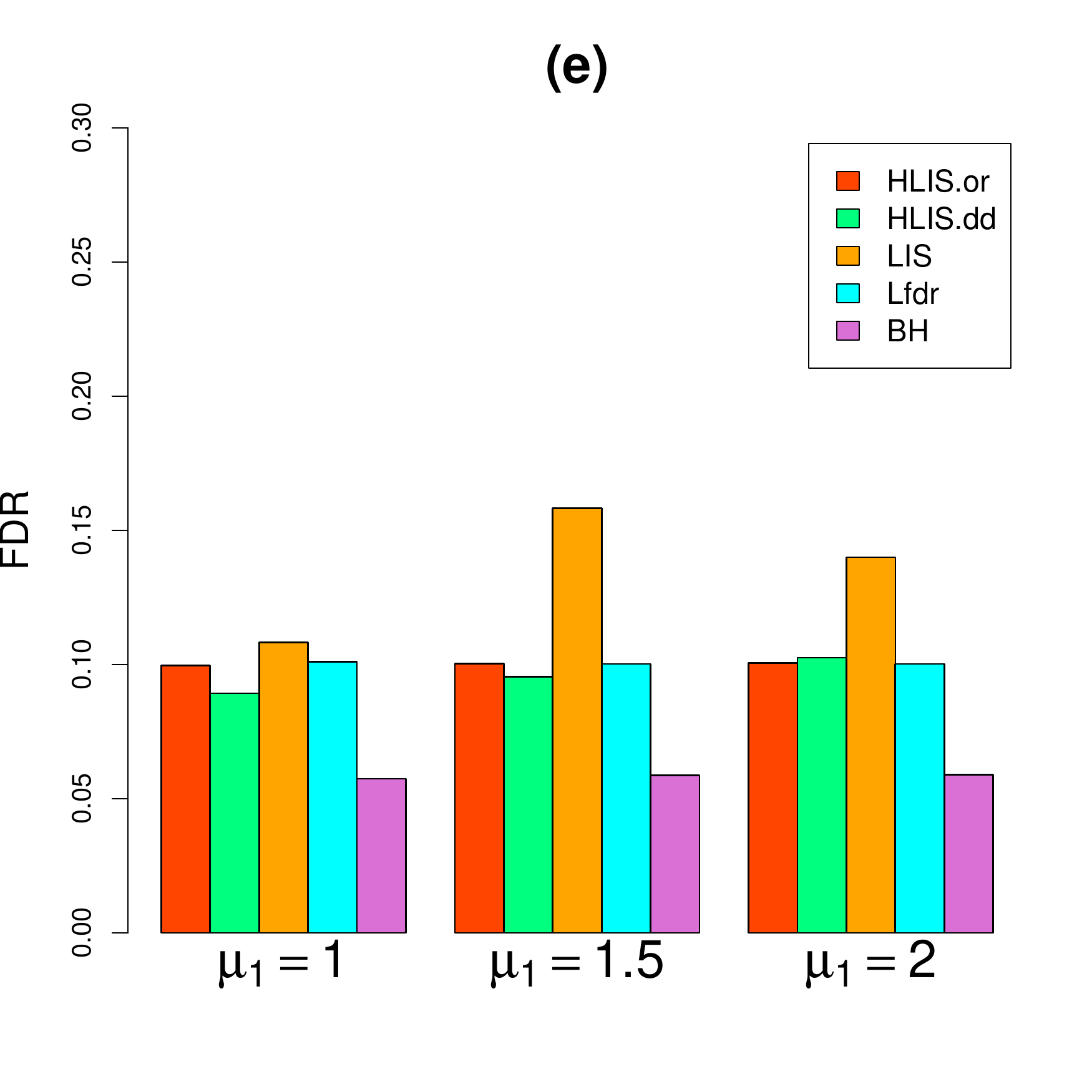}
  \end{subfigure}
  \begin{subfigure}{0.48\textwidth}
    \includegraphics[width=\textwidth,height=75mm]{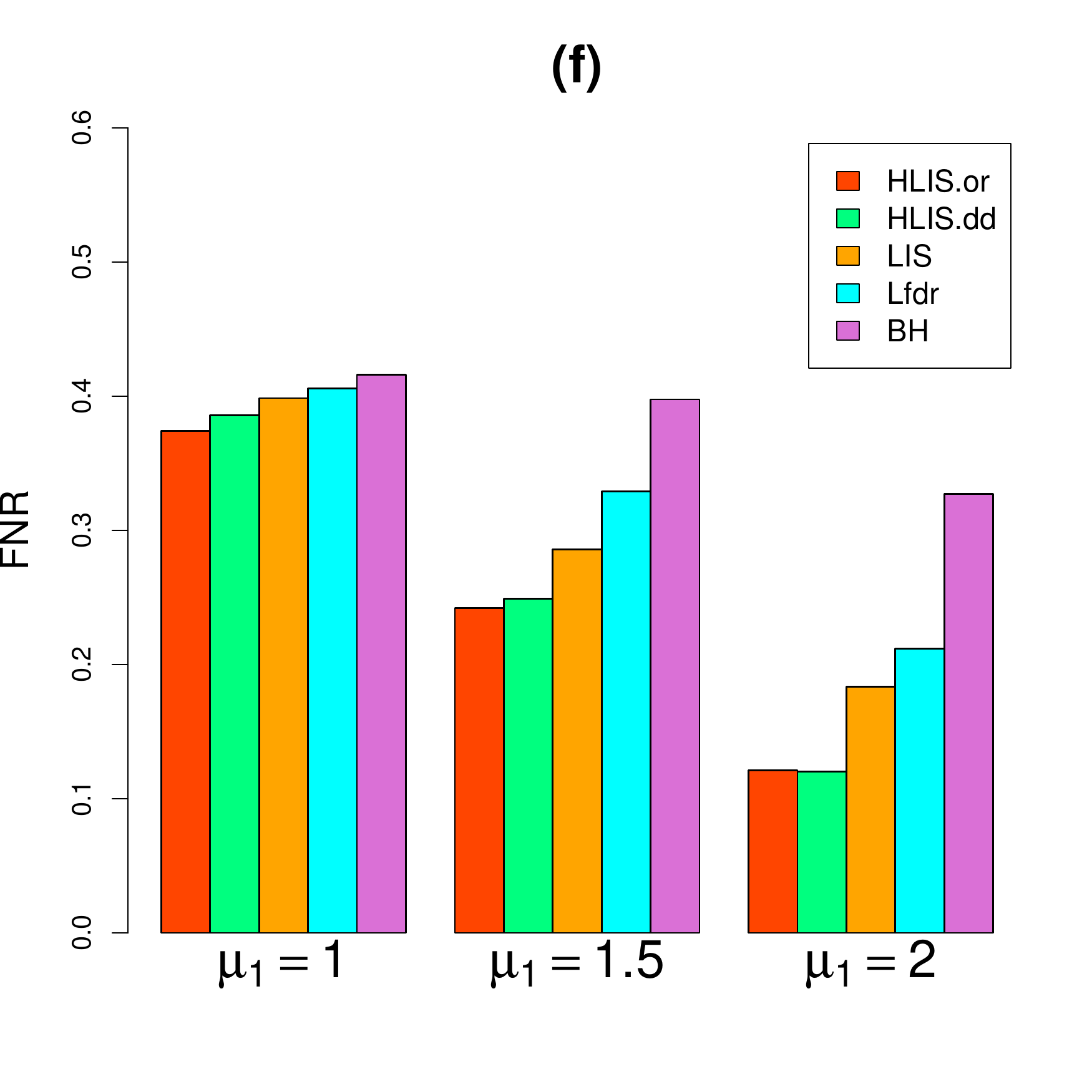}
  \end{subfigure}\\[-5mm]
  \caption{\footnotesize Simulation results in Case 1 of Simulation I: (a)-(b) simulation results in Setting 1; (c)-(d) simulation results in Setting 2; (e)-(f) simulation results in Setting 3.}
  \label{fig:2}
\end{figure}

\begin{figure}[htp]
  \centering
  \begin{subfigure}{0.48\textwidth}
    \includegraphics[width=\textwidth,height=75mm]{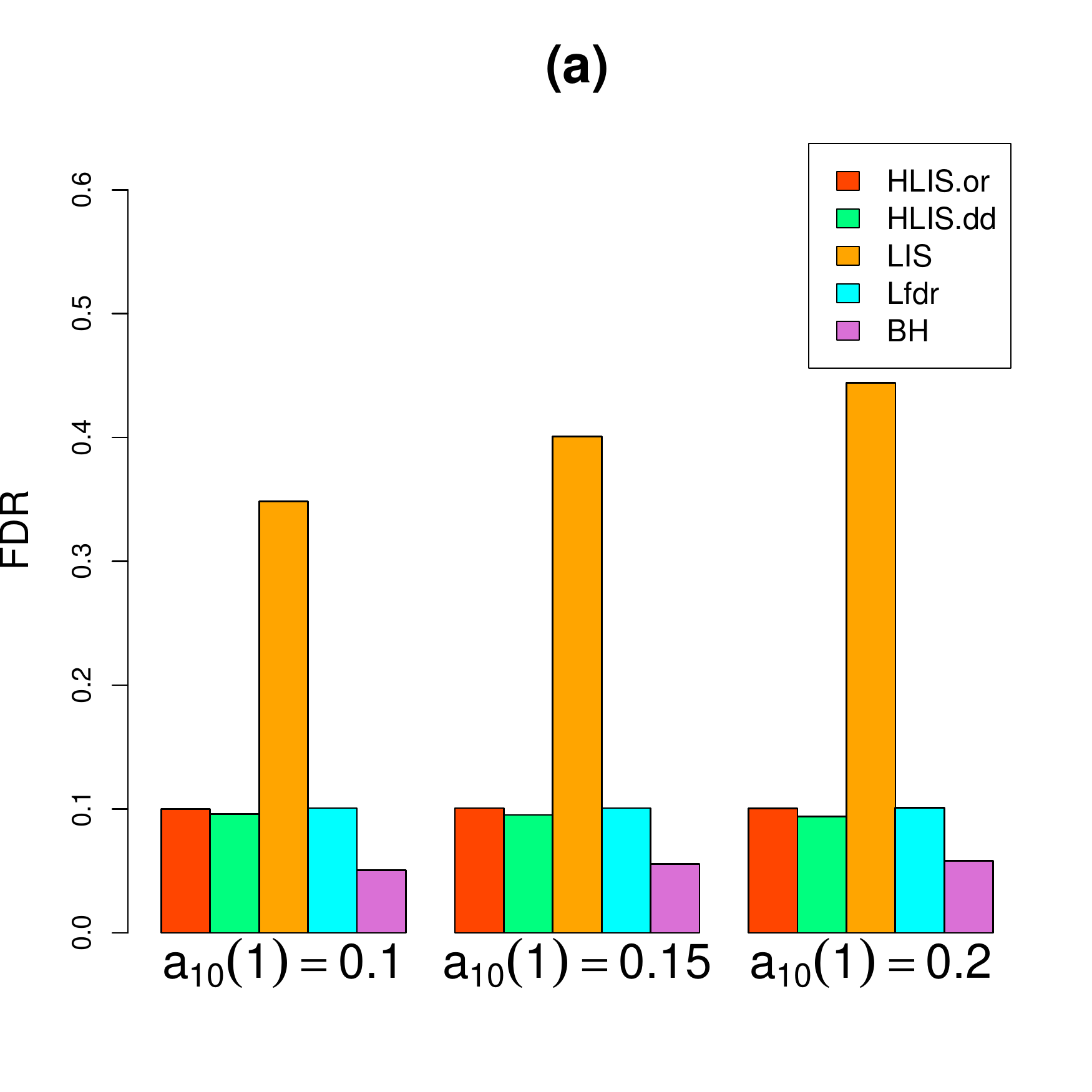}
  \end{subfigure}
  \begin{subfigure}{0.48\textwidth}
    \includegraphics[width=\textwidth,height=75mm]{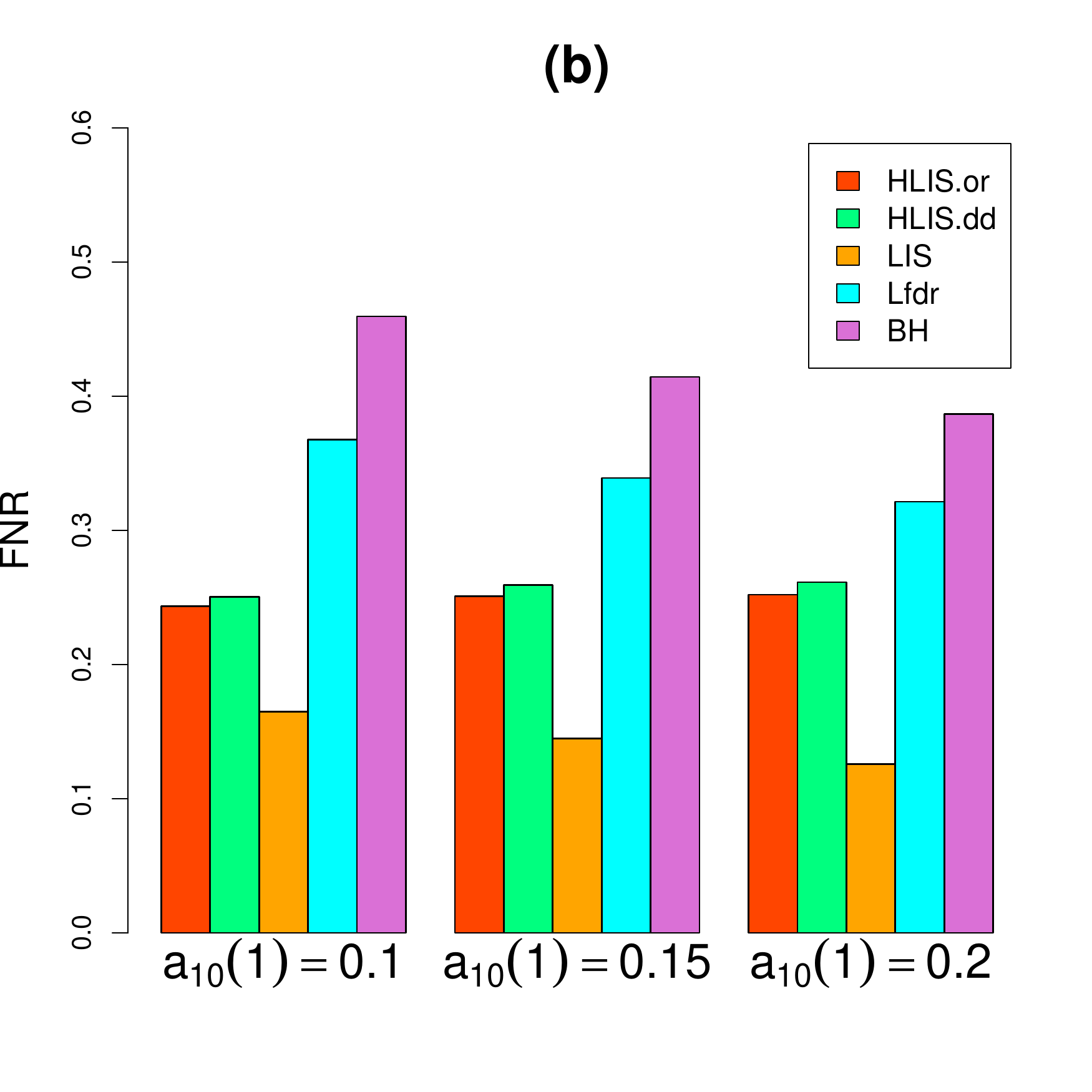}
  \end{subfigure}\\[-7mm]
  \begin{subfigure}{0.48\textwidth}
    \includegraphics[width=\textwidth,height=75mm]{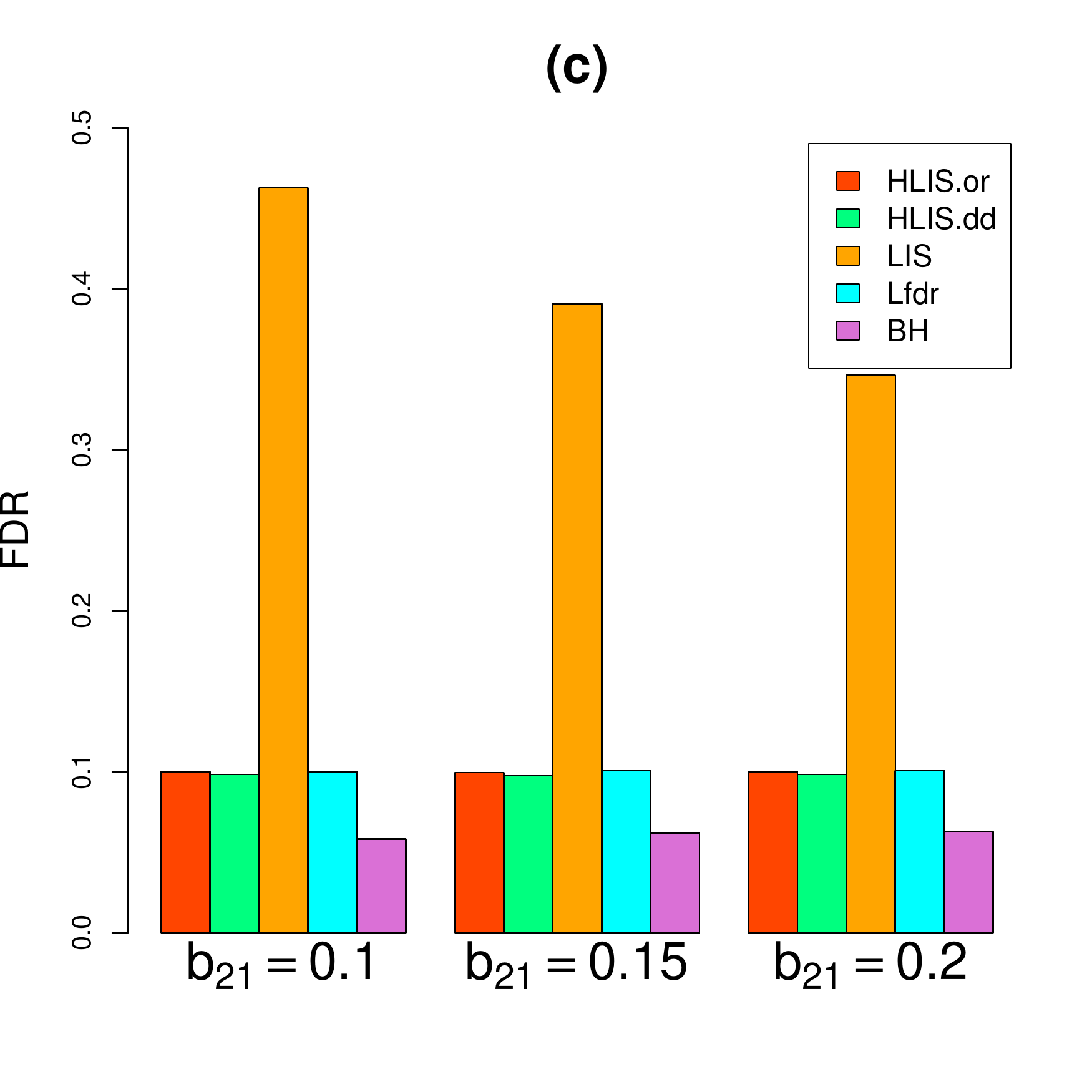}
  \end{subfigure}
  \begin{subfigure}{0.48\textwidth}
    \includegraphics[width=\textwidth,height=75mm]{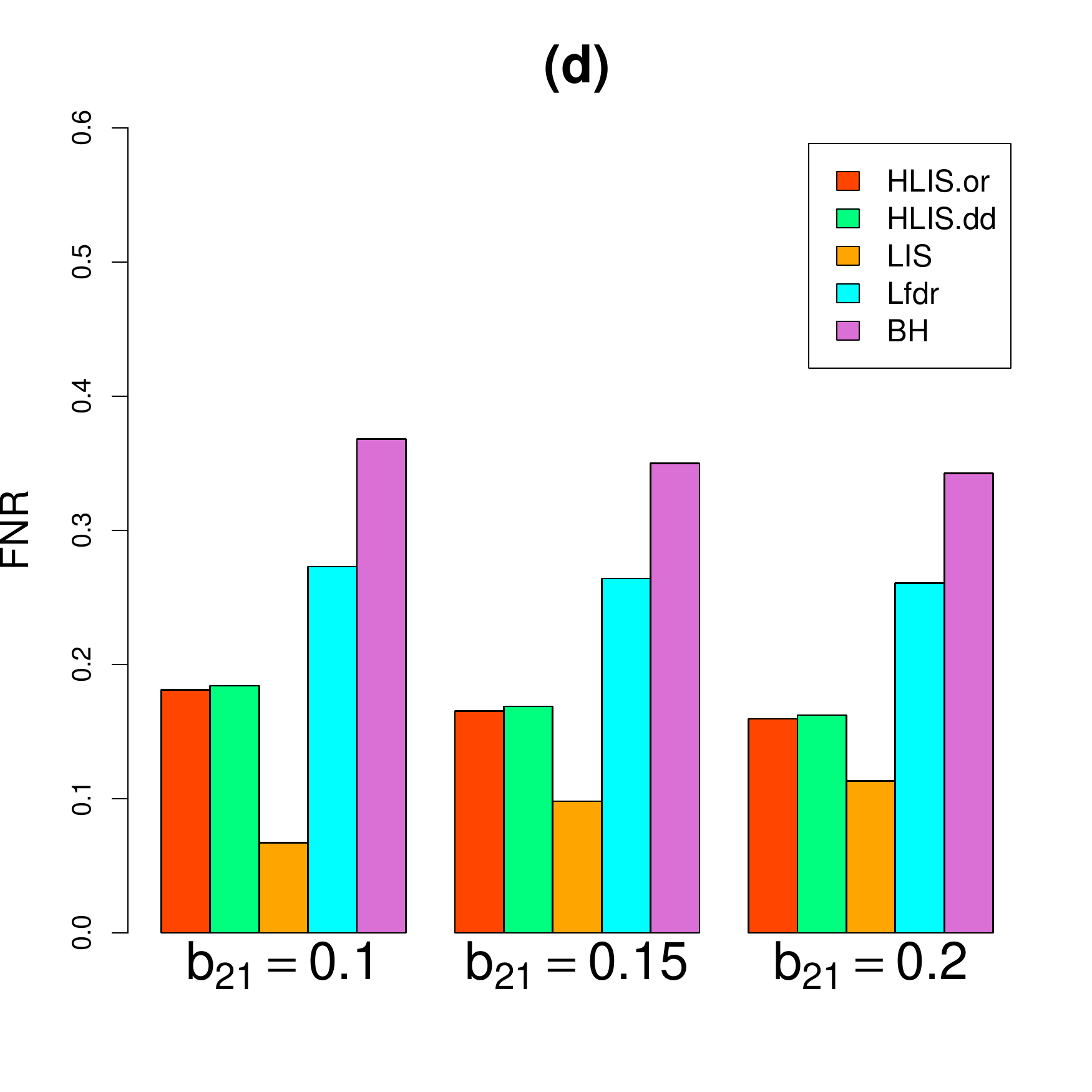}
  \end{subfigure}\\[-7mm]
  \begin{subfigure}{0.48\textwidth}
    \includegraphics[width=\textwidth,height=75mm]{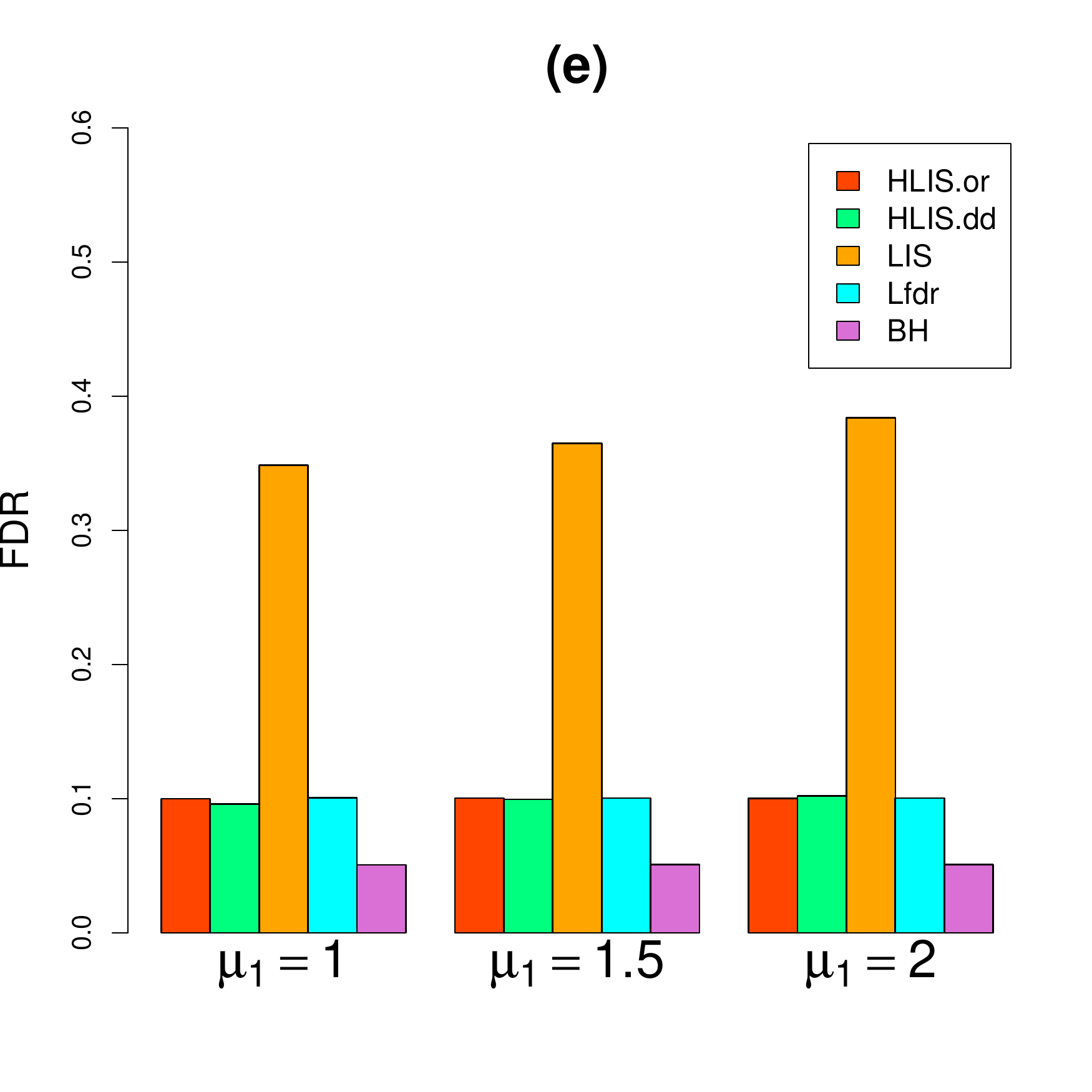}
  \end{subfigure}
  \begin{subfigure}{0.48\textwidth}
    \includegraphics[width=\textwidth,height=75mm]{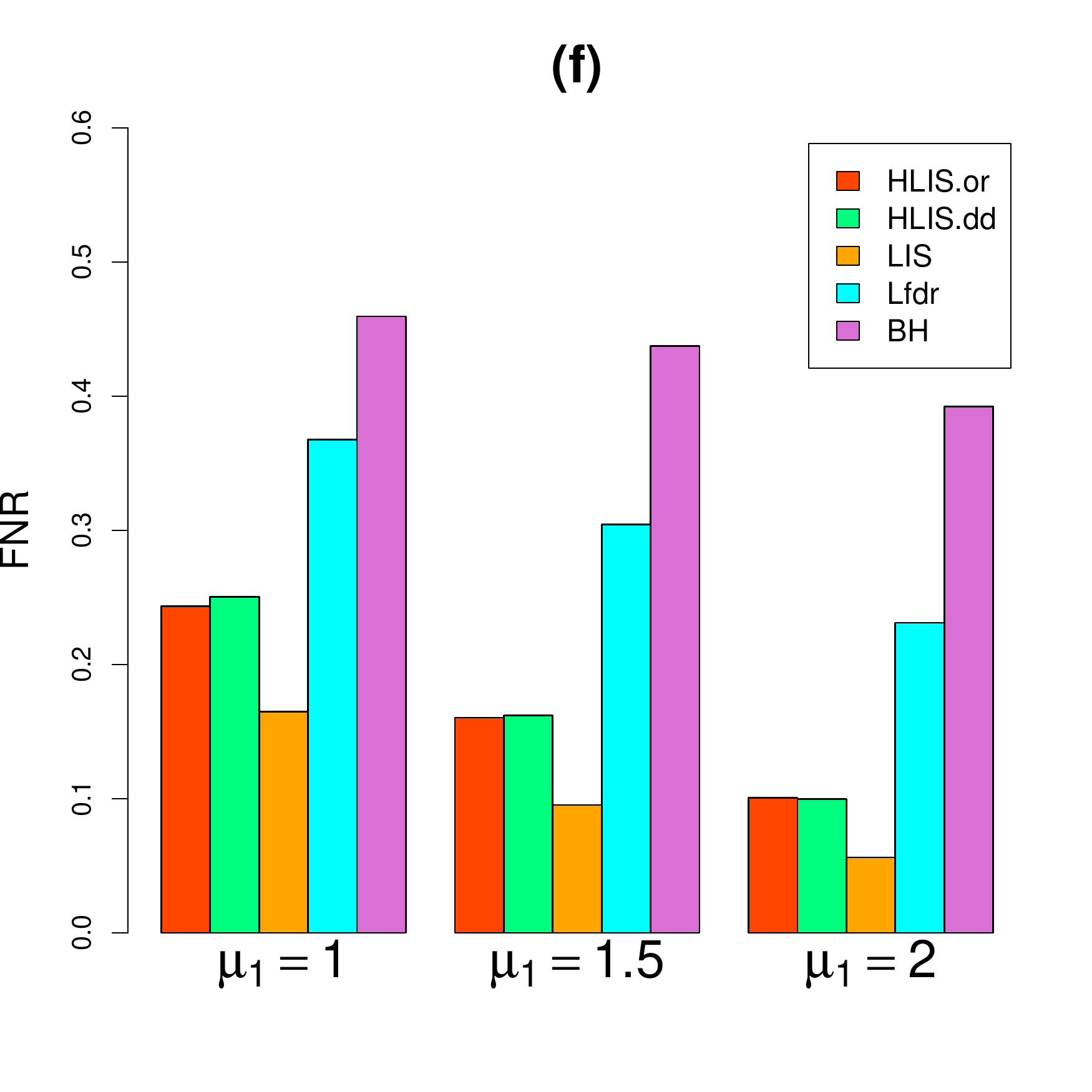}
  \end{subfigure}\\[-5mm]
  \caption{\footnotesize Simulation results in Case 1 of Simulation I: (a)-(b) simulation results in Setting 4; (c)-(d) simulation results in Setting 5; (e)-(f) simulation results in Setting 6.}
  \label{fig:3}
\end{figure}

\par
{\bf Case 2 ($K=3$):}

\par In Case 2, $\{\eta_i\}^m_{i=1}$ are generated from a block-wise Markov chain with the initial probabilities: $\boldsymbol{\pi}=(0.4, 0.3, 0.3)$, and the block-wise transition probability matrix:
\begin{equation*}       
\boldsymbol{\mathcal{B}} =
\left(                 
  \begin{array}{ccc}   
    0.8    & 0.1        & 0.1\\  
    b_{21} & 0.9-b_{21} & 0.1\\  
    0.1    & 0.1        & 0.8
  \end{array}
\right).                 
\end{equation*}
Then $\{\theta_i\}^m_{i=1}$ are generated from a process with the initial probabilities:
\begin{equation*}       
\boldsymbol{c} =
\left(                 
  \begin{array}{ccc}   
    0.5 & 0.5 & 0.5\\  
    0.5 & 0.5 & 0.5\\  
  \end{array}
\right),               
\end{equation*}
and the transition probability matrices:
\[
    \mathcal{A}_1 = \left( {\begin{array}{*{20}c} 0.9 & 0.1 \\ a_{10}(1) & 1-a_{10}(1) \\\end{array}} \right), \quad \mathcal{A}_2 = \left( {\begin{array}{*{20}c} 0.3 & 0.7 \\ 0.7 & 0.3 \\\end{array}} \right), \quad \mathcal{A}_3 = \left( {\begin{array}{*{20}c} 0.7 & 0.3 \\ 0.2 & 0.8 \\\end{array}} \right).
\]
We perform simulations under the following parameter settings.

\par
{\bf Setting 7:} fix $\lambda=1$, $\mu_1=2$, $b_{21}=0.1$ and change $a_{10}(1)$ from $0.1$ to $0.2$.

\par
{\bf Setting 8:} fix $\lambda=1$, $\mu_1=2$, $a_{10}(1)=0.2$ and change $b_{21}$ from $0.1$ to $0.2$.

\par
{\bf Setting 9:} fix $\lambda=1$, $b_{21}=0.1$, $a_{10}(1)=0.2$ and change $\mu_1$ from $1$ to $2$.

\par
{\bf Setting 10:} fix $\lambda=0.5$, $\mu_1=1$, $b_{21}=0.1$ and change $a_{10}(1)$ from $0.1$ to $0.2$.

\par
{\bf Setting 11:} fix $\lambda=0.5$, $\mu_1=1$, $a_{10}(1)=0.2$ and change $b_{21}$ from $0.1$ to $0.2$.

\par
{\bf Setting 12:} fix $\lambda=0.5$, $b_{21}=0.1$, $a_{10}(1)=0.2$ and change $\mu_1$ from $1$ to $2$.

\par
The corresponding simulation results are presented in Figures 4-5. Overall, we can get similar results as in Case 1, and hence the same conclusion can be drawn.

\begin{figure}[htp]
  \centering
  \begin{subfigure}{0.48\textwidth}
    \includegraphics[width=\textwidth,height=75mm]{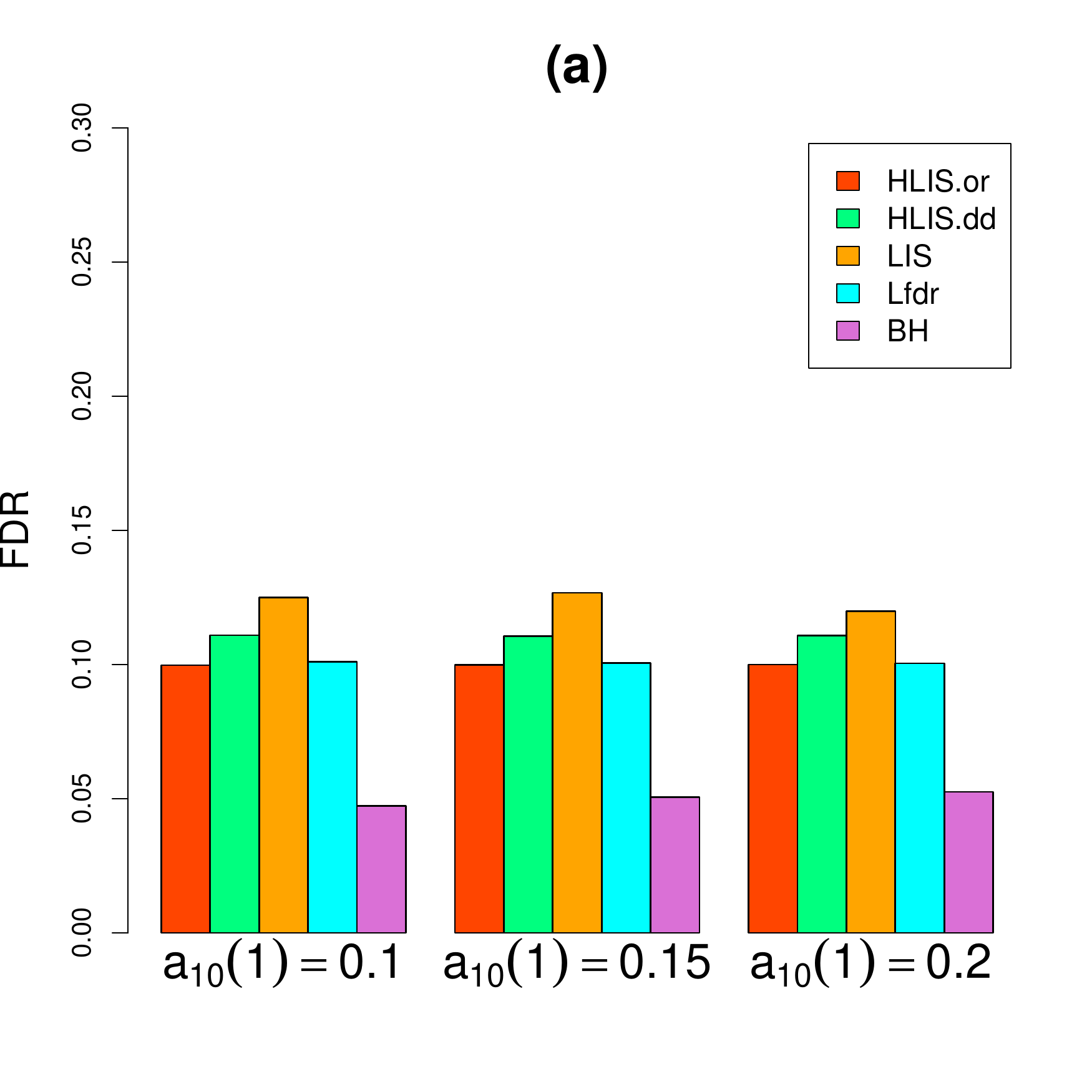}
  \end{subfigure}
  \begin{subfigure}{0.48\textwidth}
    \includegraphics[width=\textwidth,height=75mm]{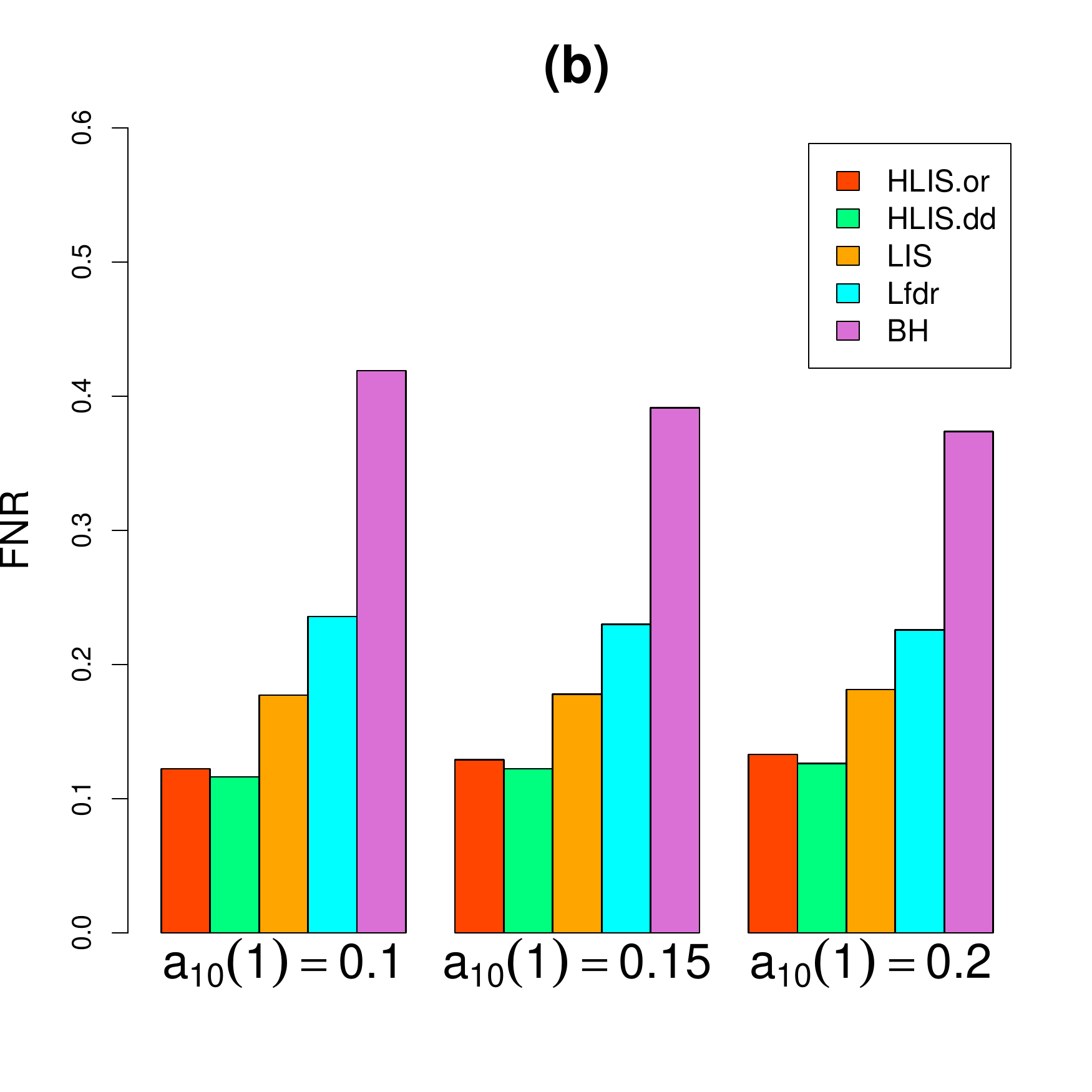}
  \end{subfigure}\\[-7mm]
  \begin{subfigure}{0.48\textwidth}
    \includegraphics[width=\textwidth,height=75mm]{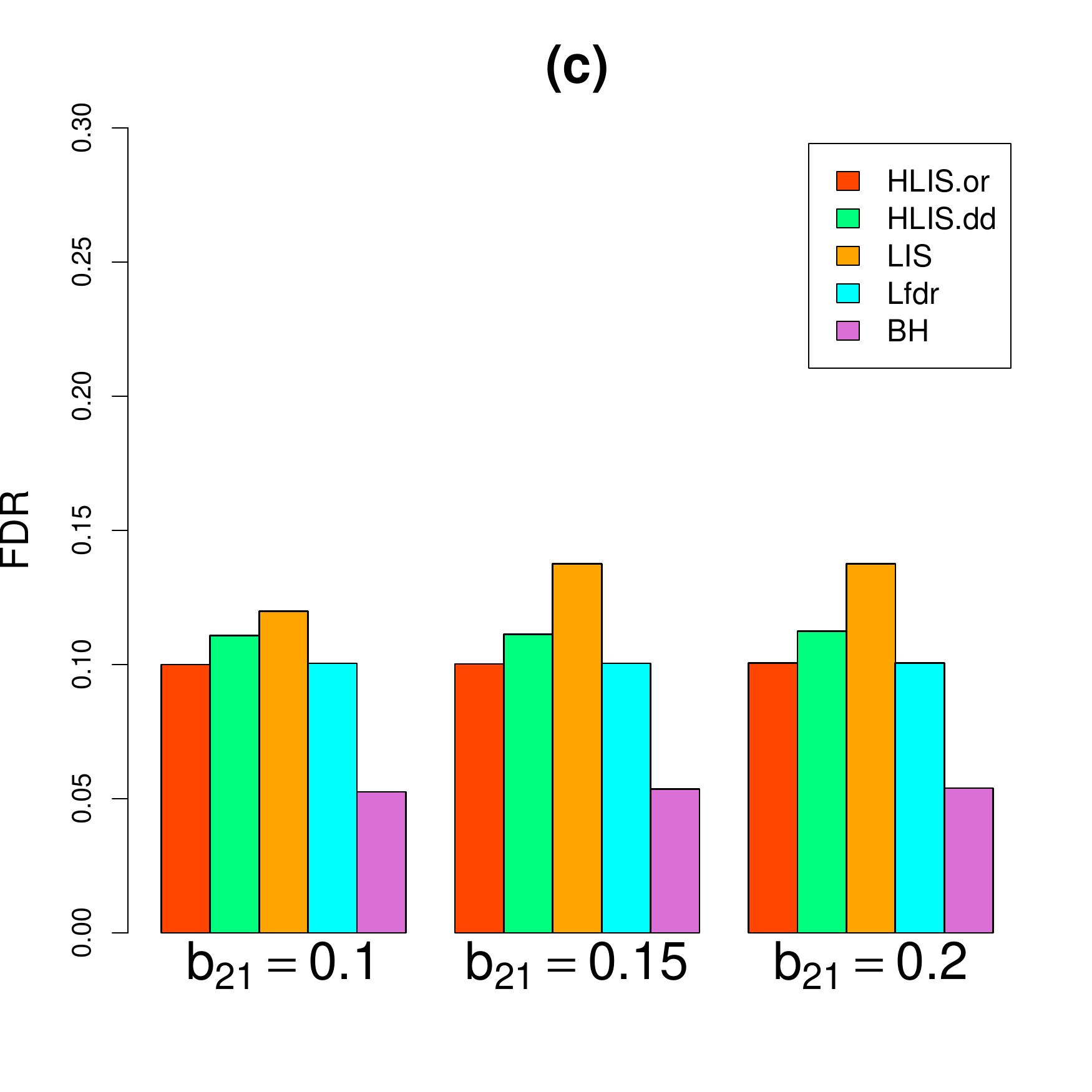}
  \end{subfigure}
  \begin{subfigure}{0.48\textwidth}
    \includegraphics[width=\textwidth,height=75mm]{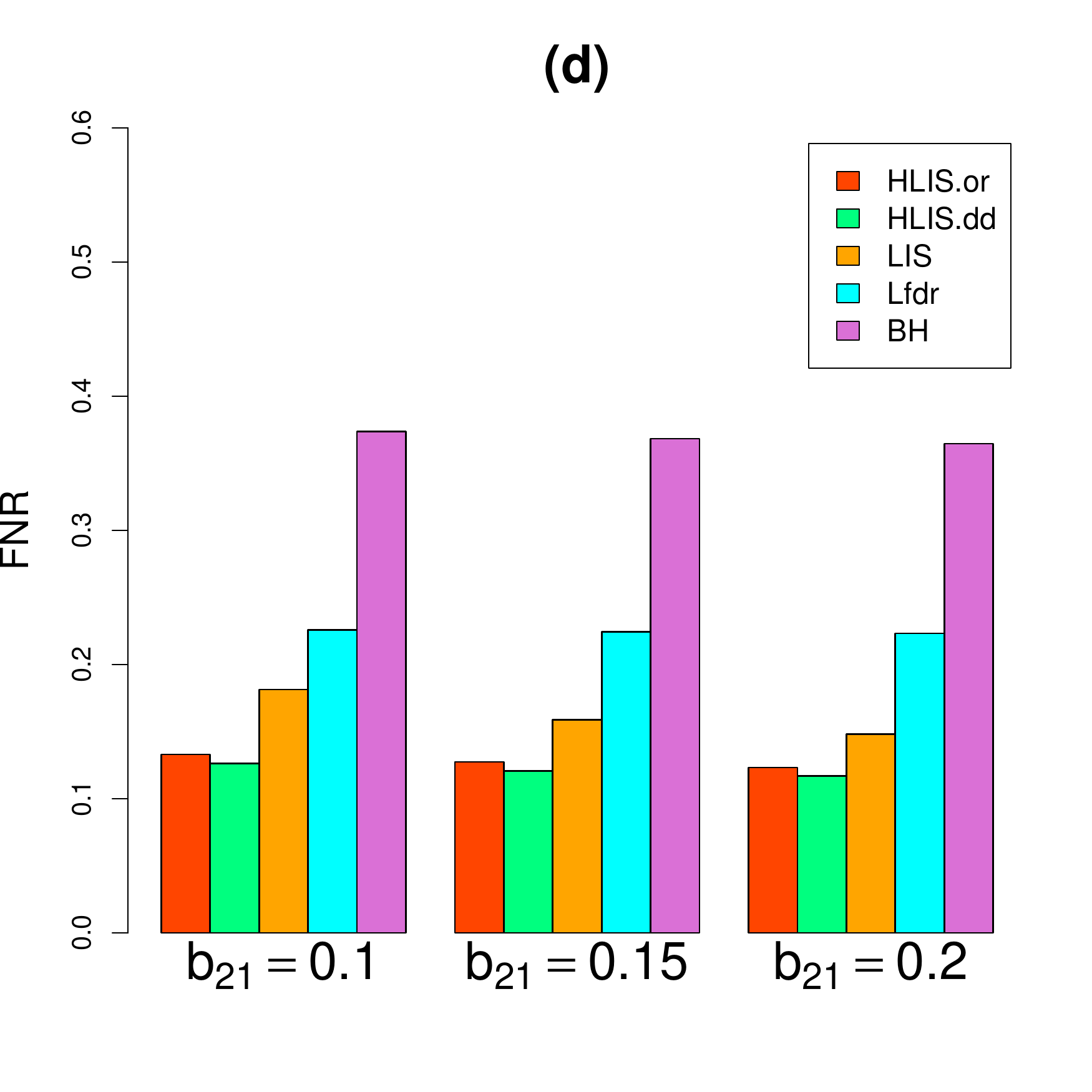}
  \end{subfigure}\\[-7mm]
  \begin{subfigure}{0.48\textwidth}
    \includegraphics[width=\textwidth,height=75mm]{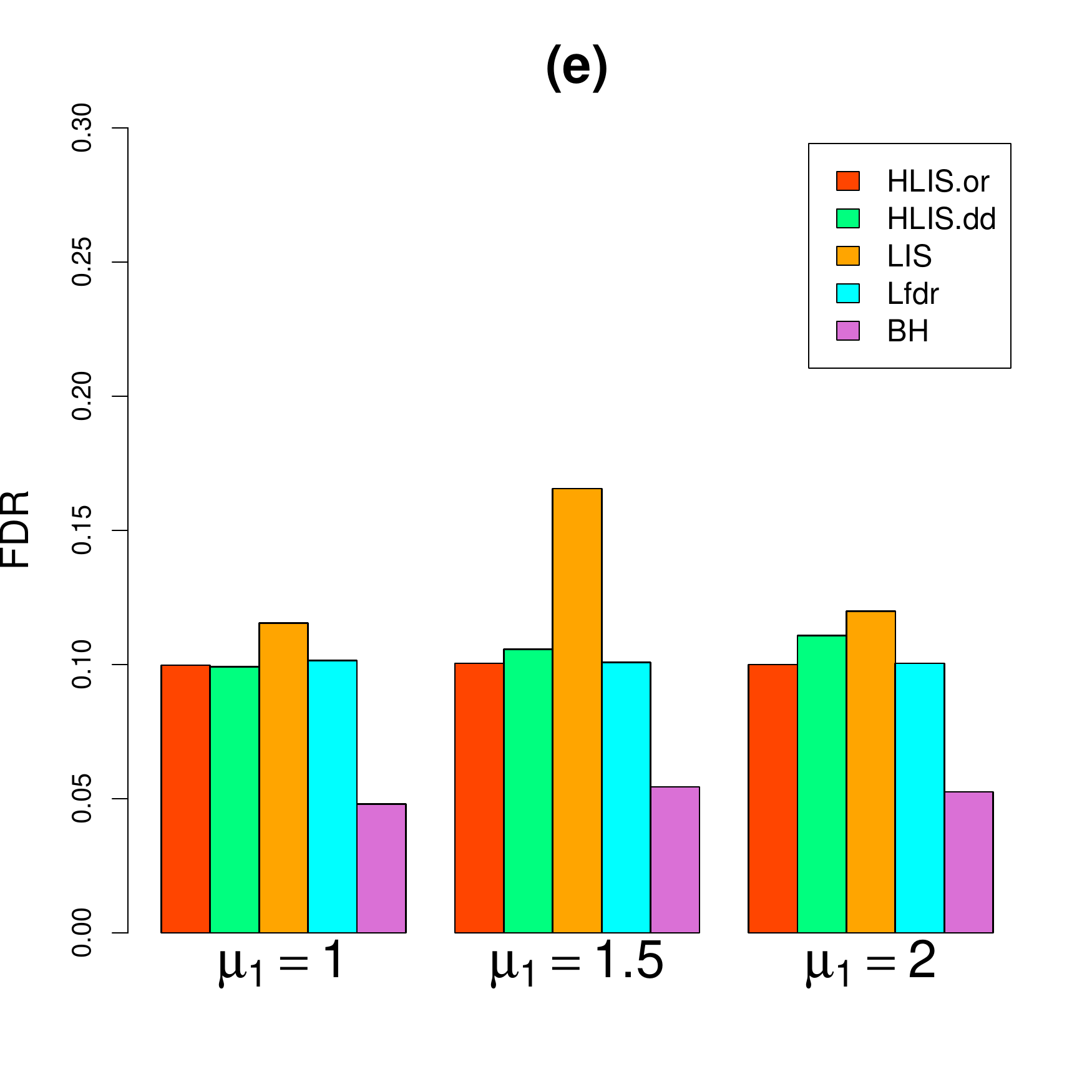}
  \end{subfigure}
  \begin{subfigure}{0.48\textwidth}
    \includegraphics[width=\textwidth,height=75mm]{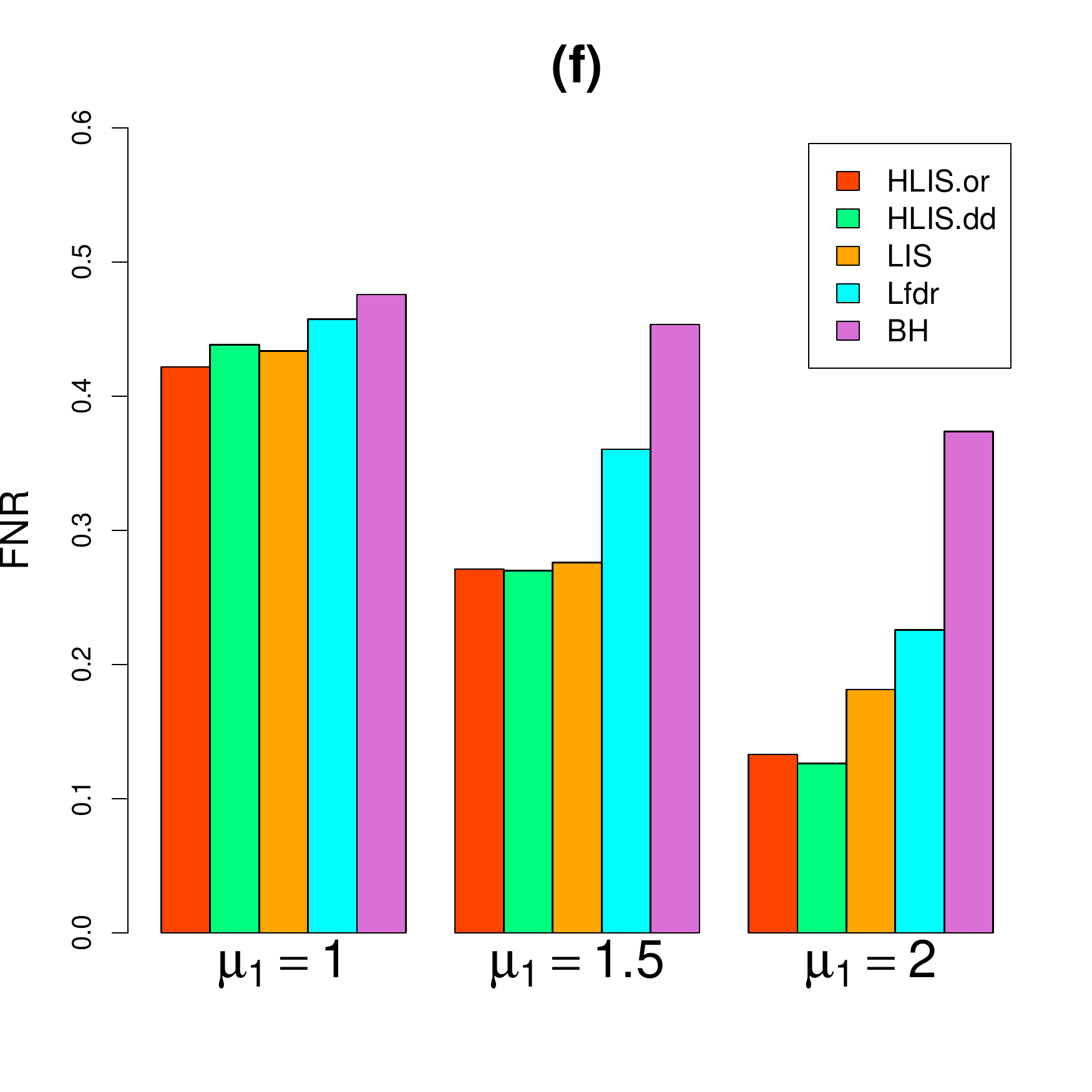}
  \end{subfigure}\\[-5mm]
  \caption{\footnotesize Simulation results in Case 2 of Simulation I: (a)-(b) simulation results in Setting 7; (c)-(d) simulation results in Setting 8; (e)-(f) simulation results in Setting 9.}
  \label{fig:4}
\end{figure}

\begin{figure}[htp]
  \centering
  \begin{subfigure}{0.48\textwidth}
    \includegraphics[width=\textwidth,height=75mm]{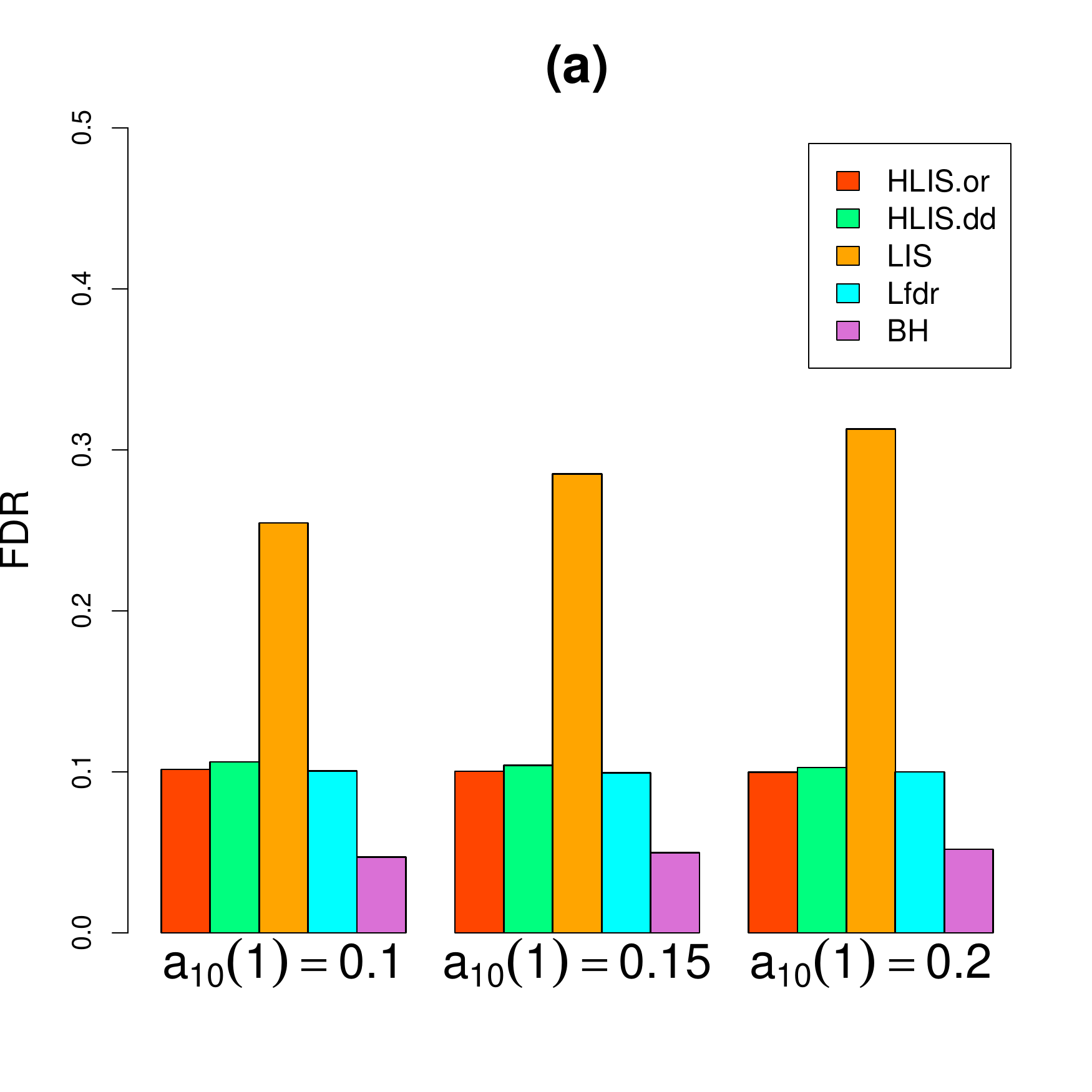}
  \end{subfigure}
  \begin{subfigure}{0.48\textwidth}
    \includegraphics[width=\textwidth,height=75mm]{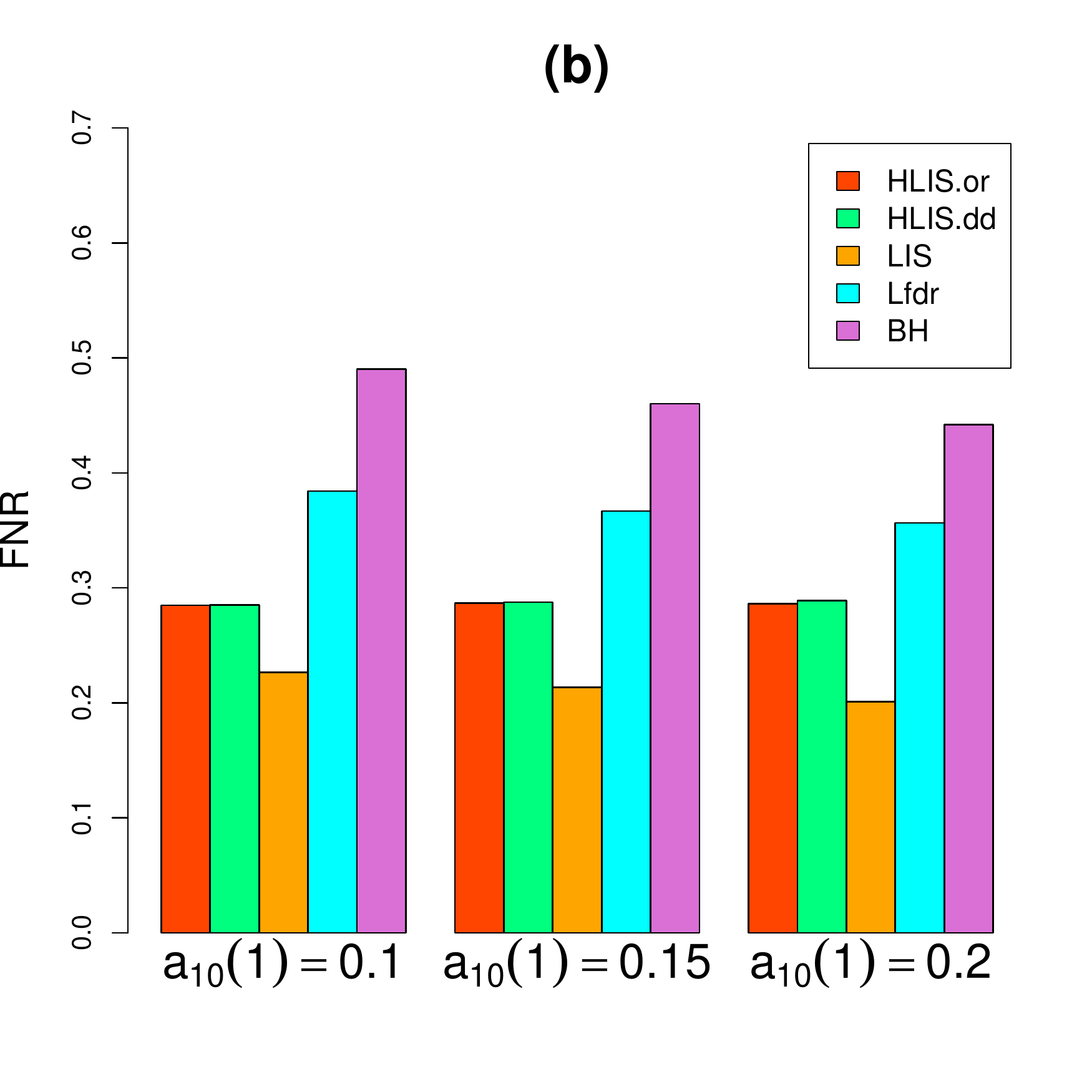}
  \end{subfigure}\\[-7mm]
  \begin{subfigure}{0.48\textwidth}
    \includegraphics[width=\textwidth,height=75mm]{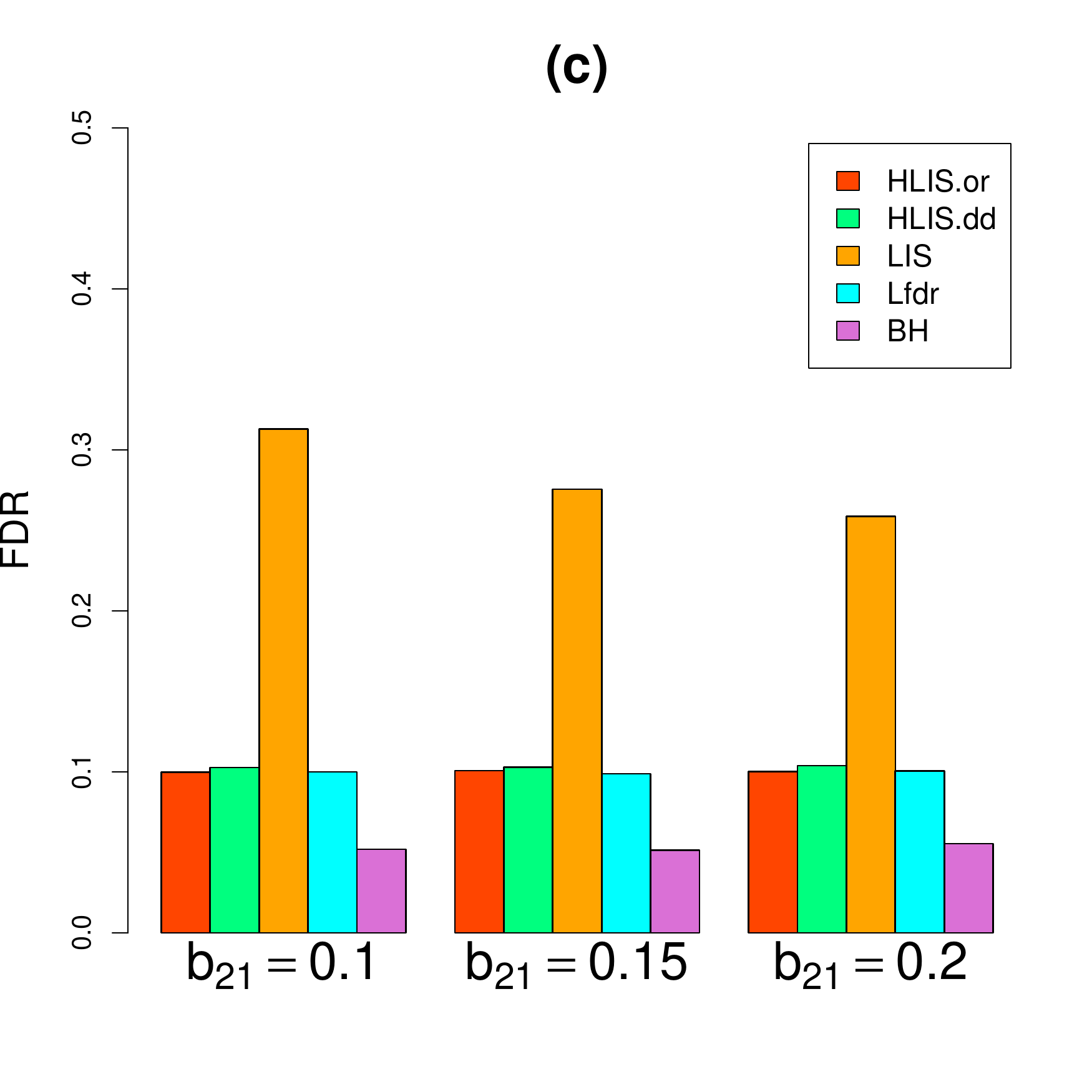}
  \end{subfigure}
  \begin{subfigure}{0.48\textwidth}
    \includegraphics[width=\textwidth,height=75mm]{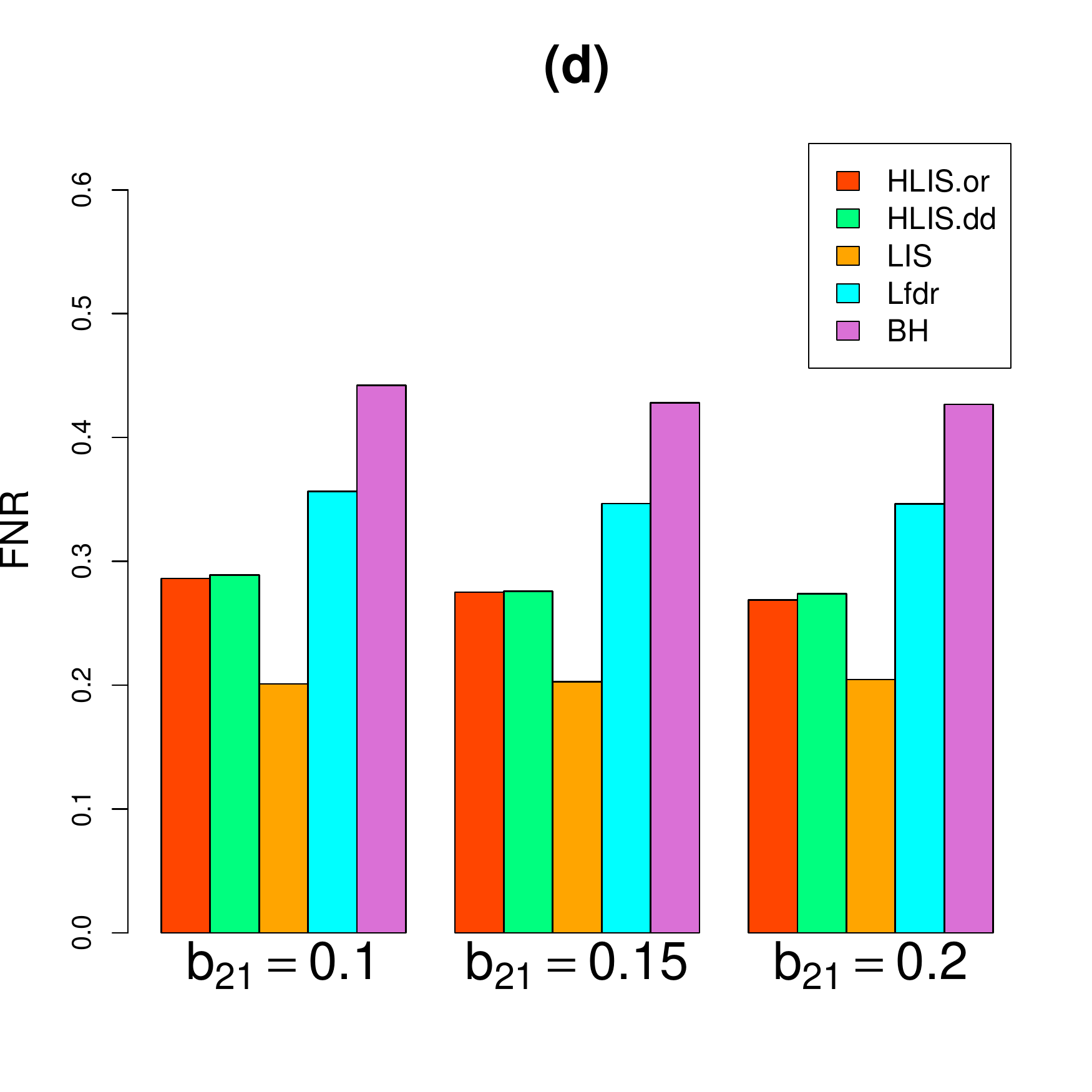}
  \end{subfigure}\\[-7mm]
  \begin{subfigure}{0.48\textwidth}
    \includegraphics[width=\textwidth,height=75mm]{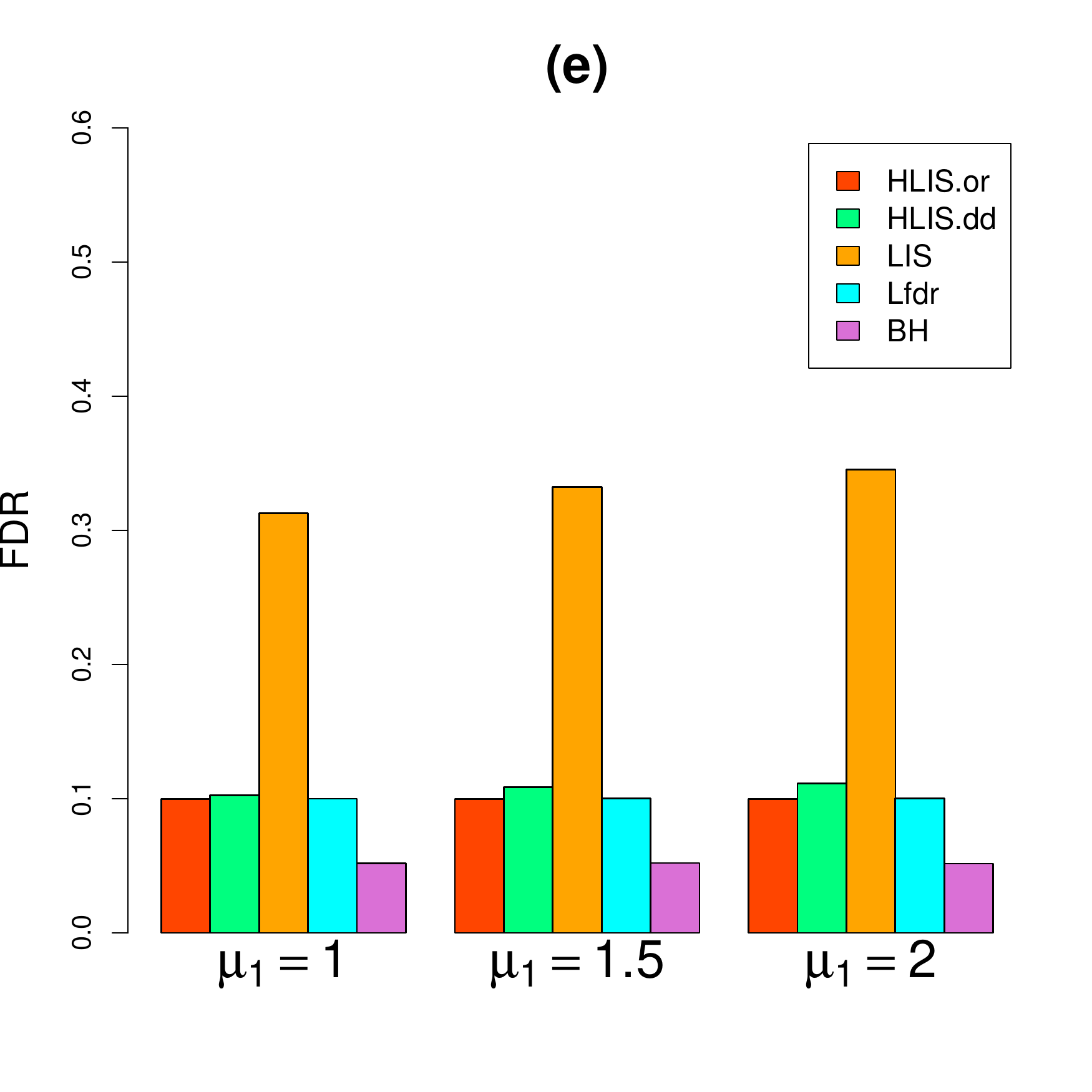}
  \end{subfigure}
  \begin{subfigure}{0.48\textwidth}
    \includegraphics[width=\textwidth,height=75mm]{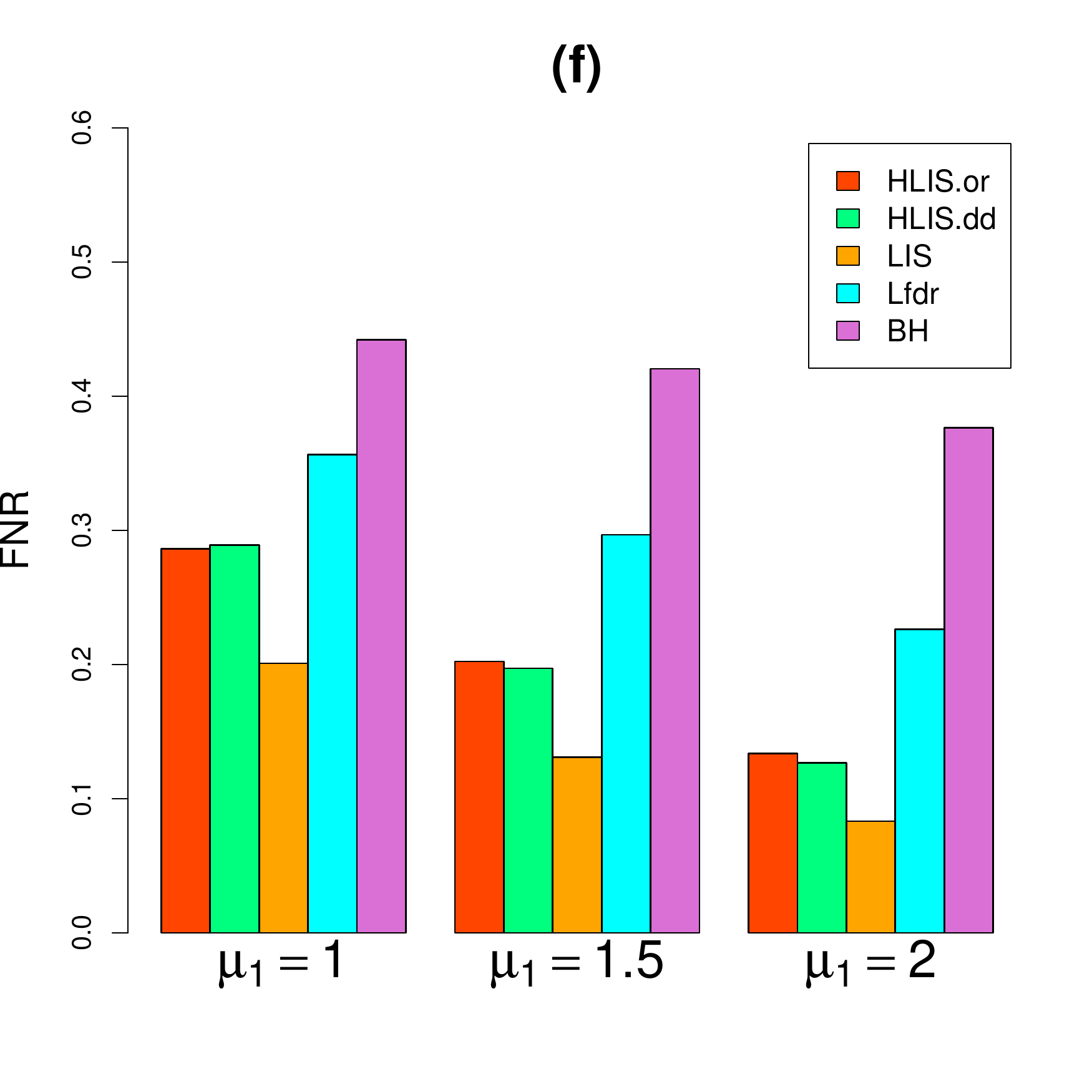}
  \end{subfigure}\\[-5mm]
  \caption{\footnotesize Simulation results in Case 2 of Simulation I: (a)-(b) simulation results in Setting 10; (c)-(d) simulation results in Setting 11; (e)-(f) simulation results in Setting 12.}
  \label{fig:5}
\end{figure}

\subsection{Simulation II}\label{sec-3.2}

\par
In Simulation II, we examine the numerical performance of the HLIS procedure on a more realistic simulated data. Specifically, we generate a genotype pool by randomly matching $340$ haplotypes from the subjects of JPT+CHB (Japanese in Tokyo, Japan and Han Chinese in Beijing, China) collected by HapMap3 \citep{HapMap3}. Without loss of generality, eight SNPs from a region of chromosome $7$ (9000 SNPs in total) are selected as disease- or trait-related SNPs, of which four selected SNPs (the 2000th, 2500th, 3000th and 3500th) are distant from each other and four SNPs (the 6000th, 6010th, 6020th and 6030th) are close together. Intuitively, there are three regions of chromosomes (2000-3500, 6000-6030, others). Thus the HLIS procedure is conducted with $K=3$ in Simulation II. The disease or trait status $Y$ is generated by the logistic regression model:
\[
        \log \frac{\Pr(Y=1 \mid \boldsymbol{G})}{1-\Pr(Y=1 \mid \boldsymbol{G})} = \beta_0 + \sum_{i=1}^{8} \beta_i G_i,
\]
where $\boldsymbol{G}=(G_1, G_2, \cdots, G_8)$ and $G_i=0, 1, 2$ is the genotype with respect to the $i$th SNP. Let $\beta_0=-5$ and $\beta_1=\beta_2=\cdots=\beta_8=\log(2)$, where the prevalence is thus approximated by $0.03$. The numerical performance of these multiple testing procedures is assessed by the selection rate of the associated SNPs under different top $k$ SNPs, where the associated SNPs are referred to as the five adjacent SNPs on each side of the causal SNP. The corresponding simulation results are displayed in Figure 6. We can see from Figure 6 that the selection rate yielded by the HLIS procedure is consistently larger than those of other procedures. This illustrates that the HLIS procedure has a higher ranking efficiency.

\begin{figure}[H]
\centering
\includegraphics[height=4.8in,width=5in]{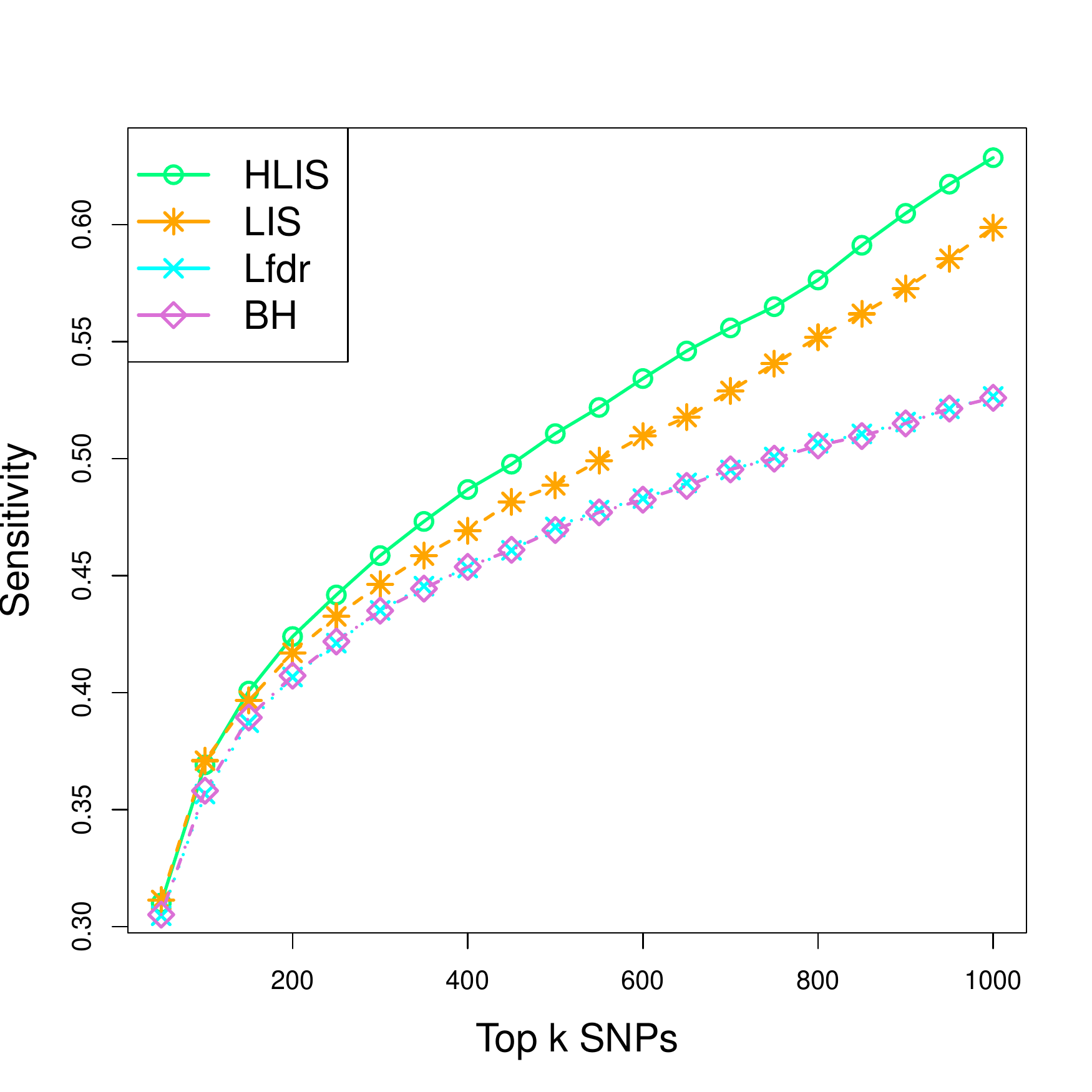}\\[-5mm]
\caption{\footnotesize{The sensitivity curves}}
\end{figure}

\section{Real Data Analysis}

\par
Schizophrenia (SCZ) is a heritable disorder that has significant public health implications. \cite{lichtenstein2006recurrence} reported that genetic variation plays an important role in the etiology of SCZ. To demonstrate the effectiveness of the HLIS procedure in practical applications, we apply the HLIS procedure to detect SNPs associated with SCZ. The corresponding data is collected by the Psychiatric Genetics Consortium (PGC) and available publicly on the websites https://www.med.unc.edu/pgc/download-results/scz/. The SCZ data consists of a meta-analysis of 5001 SCZ cases and 6243 controls from the Swedish samples, and 8832 SCZ cases and 12067 controls from independent PGC SCZ samples \citep{ripke2013genome}. The statistic is calculated by the following formula
\[
    z_i = \log (\text{OR}_i)/\text{SE}_i,
\]
where $\text{OR}_i$ is the odds ratio for the $i$th effect allele and $\text{SE}_i$ is the standard error of $\log (\text{OR}_i)$. For the illustrative purpose only, we restrict attention to detect SNPs associated with SCZ on Chromosome 22. Note that the number of chromosomal region types $K$ and the number of mixed components for the non-null $L$ are unknown in practice. To deal with this issue, we use the Bayesian information criterion (BIC) to select $K$ and $L$ for the HLIS procedure and the LIS procedure, respectively. By comparing the BIC values, both $K$ and $L$ are chosen to be $2$.

\par
The detailed results are listed in Figure 7. This figure displays the number of discoveries identified by different procedures relative to the target FDR level varied from $0$ to $2\times 10^{-6}$. We can see from Figure 7 that the HLIS procedure identifies more SNPs associated with SCZ at various FDR levels. The poor performance of the Lfdr procedure may be due to the inaccurate estimation of $\Pr(\theta_i=0)$ as a result of not taking into account local correlations. By and large, these results demonstrate that the HLIS procedure is more efficient by leveraging the HHMM.

\begin{figure}[H]
\centering
\includegraphics[height=4.8in,width=5in]{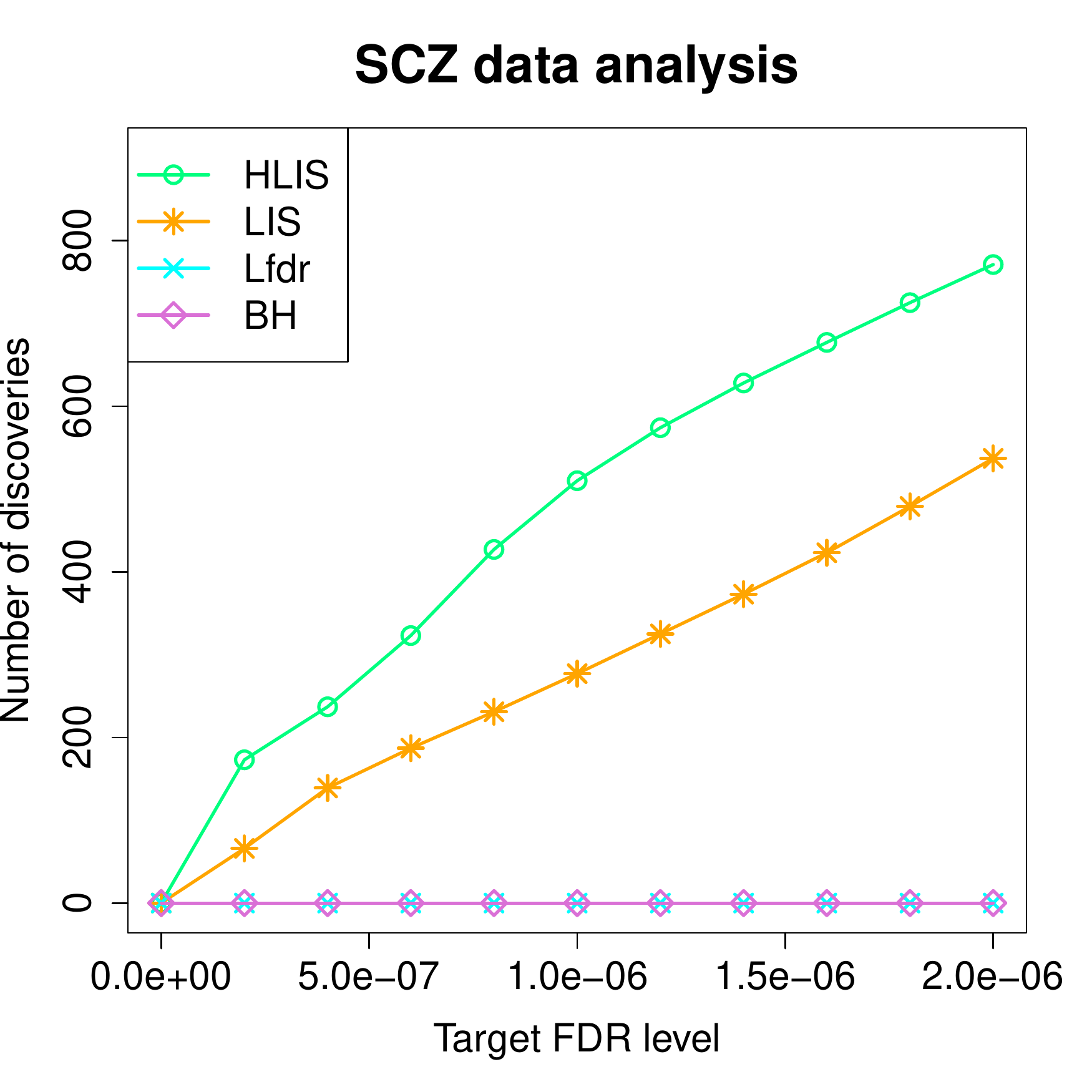}\\[-5mm]
\caption{\footnotesize{The number of discoveries relative to the target FDR level in SCZ data analysis.}}
\end{figure}

\section{Discussion}

\par
This paper develops a novel multiple testing procedure based on the HHMM for GWAS. The HLIS procedure can automatically divide the different types of chromosome regions and also takes into account local correlations among tests. In essence, the HLIS statistic can be viewed as a weighted LIS statistic. Theoretically, it is shown that the oracle HLIS procedure is valid and optimal in some sense. Then a data-driven HLIS procedure is proposed to mimic the oracle version. Extensive simulations and the real data analysis illustrate the effectiveness of the HLIS procedure. Although the power of the HLIS procedure has been significantly improved, it can be extended in several ways.

\par
First, in practice, the $z$-values or the statistics may be influenced by the covariates. For example, it has been shown that the etiology of complex disease depends not only on the genetic effects but also on the covariates \citep{zhu2012nonparametric}. A proper use of covariate-adjustment in GWAS not only improves the efficacy of multiple testing but also increases the interpretability of the results. One way of extending the HLIS procedure to take into account covariate effects is to drop the assumption that the Markov chain is homogeneous. Such an extension of the LIS procedure can be found in \cite{kuan2012integrating}.

\par
Second, the EM algorithm for estimating HHMM parameters is a heuristic algorithm. Note that the MLE obtained by the EM algorithm is only the local maximum of the likelihood function. An inappropriate choice of initial values may lead to poor parameter estimations. An alternative algorithm to bypass this issue is to employ Bayesian sampling algorithm for parameter estimation. This extension is outside of the scope of this paper.

\par
Third, recently, \cite{Denti2021Two} suggested to employ mixtures of two-parameter Poisson-Dirichlet (2PPD) processes instead of the two-component mixture model for multiple testing. They demonstrated that the 2PPD processes provide a more flexible and effective tool for large-scale hypothesis tesing. Such an extension of the HLIS procedure based on the 2PPD processes is meaningful and challenging. We plan to leave the exploration of this issue for our future research.

\newpage
\section{Appendix}

\par
{\bf Proof of Theorem 1}
\begin{proof}
By the continuity of the pdf of $\{Z_i\}^m_{i=1}$, we have that the pdf and cdf of $\HLIS_j\left(\boldsymbol{Z}\right)$ are also continuous. Since $\mFDR({\bm\delta}(\HLIS(\boldsymbol{Z}),c))$ can be expressed as
\[
    \mFDR\left({\bm\delta}(\HLIS(\boldsymbol{Z}), c)\right)=\dfrac{\sum\limits^m_{j=1}\Pr(\HLIS_j(\boldsymbol{Z})<c, \theta_j=0\mid {\boldsymbol{\vartheta}})}{\sum\limits^m_{j=1}\Pr(\HLIS_j(\boldsymbol{Z})<c)\mid {\boldsymbol{\vartheta}})},
\]
$\mFDR({\bm\delta}(\HLIS(\boldsymbol{Z}),c))$ is continuous with respect to $c$.

\par
Let $\Lambda_j(\boldsymbol{Z})=\HLIS_j\left(\boldsymbol{Z}\right)/(1-\HLIS_j\left(\boldsymbol{Z}\right))$, for $j=1,\cdots, m$. It can be shown that $\Lambda_j(\boldsymbol{Z})$ satisfies the monotone ratio condition (MRC) defined in \cite{sun2009large}. By Theorem 1 of \cite{sun2009large}, we can conclude that $\mFDR({\bm\delta}(\Lambda(\boldsymbol{Z}),c))$ is strictly increasing in $c$. Note that ${\bm\delta}(\HLIS(\boldsymbol{Z}), c)={\bm\delta}(\Lambda(\boldsymbol{Z}), \varphi(c))$, where $\varphi(c)=c/(1-c)$ is strictly increasing in $c$. Then, we have that $\mFDR({\bm\delta}(\HLIS(\boldsymbol{Z}),c))$ is strictly increasing in $c$.

\par
Moreover, taking the limits for $\mFDR({\bm\delta}(\HLIS(\boldsymbol{Z}),c))$, we have that
\[
    \lim\limits_{c \to 0}\mathrm{mFDR}\left({\bm\delta}(\HLIS(\boldsymbol{Z}), c)\right)=0,
\]
and
\[
    \lim\limits_{c \to 1}\mathrm{mFDR}\left({\bm\delta}(\HLIS(\boldsymbol{Z}), c)\right)=1,
\]
It follows that the set $\{t:\mathrm{mFDR}({\bm\delta}(\HLIS(\boldsymbol{Z}),t))\leq\alpha\}$ is nonempty, for any $0<\alpha<1$. Let
\[
    c_{\alpha} = \sup\{t: \mathrm{mFDR}({\bm\delta}(\HLIS(\boldsymbol{Z}),t))\leq\alpha\}.
\]
This yields that $$\mFDR({\bm\delta}(\HLIS(\boldsymbol{Z}),c_{\alpha}))=\alpha.$$
\end{proof}

\par
\leftline{\bf Proof of Theorem 2}
\begin{proof}
Let $c_{\alpha}$ be the cut-off satisfies the condition $\mFDR\left({\bm\delta}(\HLIS(\boldsymbol{Z}), c_{\alpha})\right)=\alpha$. Note that ${\bm\delta}(\HLIS(\boldsymbol{Z}), c_{\alpha})={\bm\delta}(\Lambda(\boldsymbol{Z}), \varphi(c_{\alpha}))$, where $\varphi(c_{\alpha})=c_{\alpha}/(1-c_{\alpha})$. Therefore,
\[
        \frac{\mathrm{E}\left[\sum\limits^m_{j=1}I(\Lambda_j(\boldsymbol{Z})<\varphi(c_{\alpha}))\Pr(\theta_j=0\mid\boldsymbol{Z}, \boldsymbol{\vartheta})\right]}
        {\mathrm{E}\left[\sum\limits^m_{j=1}I(\Lambda_j(\boldsymbol{Z})<\varphi(c_{\alpha}))\left(\Pr
        (\theta_j=0\mid\boldsymbol{Z}, {\boldsymbol{\vartheta}})+\Pr(\theta_j=1\mid\boldsymbol{Z}, {\boldsymbol{\vartheta}})\right)\right]}=\alpha,
\]
implying that
\[
        \sum\limits^m_{j=1}\mathrm{E}\left\{I(\Lambda_j(\boldsymbol{Z})<\varphi(c_{\alpha}))\left[\Pr(\theta_j=0\mid\boldsymbol{Z}, {\boldsymbol{\vartheta}})-\varphi(\alpha)\Pr(\theta_j=1\mid\boldsymbol{Z}, {\boldsymbol{\vartheta}})\right]\right\}=0.\eqno{(\mathrm{A}.1)}
\]
Similarly, the condition $\mFDR({\bm\delta}(T(\boldsymbol{Z}),c))\leq\alpha$ yields that
\[
        \sum\limits^m_{j=1}\mathrm{E}\left\{I(T_j(\boldsymbol{Z})<c)\left[\Pr(\theta_j=0\mid\boldsymbol{Z}, {\boldsymbol{\vartheta}})-\varphi(\alpha)\Pr(\theta_j=1\mid\boldsymbol{Z}, {\boldsymbol{\vartheta}})\right]\right\}\leq0.\eqno{(\mathrm{A}.2)}
\]
Combining (A.1) with (A.2), we can obtain that
\[
        \sum\limits^m_{j=1}\mathrm{E}\left\{[I(\Lambda_j(\boldsymbol{Z})<\varphi(c_{\alpha}))-I(T_j(\boldsymbol{Z})<c)]
        \left[\Pr(\theta_j=0\mid\boldsymbol{Z}, {\boldsymbol{\vartheta}})-\varphi(\alpha)\Pr(\theta_j=1\mid\boldsymbol{Z}, {\boldsymbol{\vartheta}})\right]\right\}\geq0.\eqno{(\mathrm{A}.3)}
\]
By the definition of $\Lambda_j(\boldsymbol{Z})$, we have that
\[
        \sum\limits^m_{j=1}\mathrm{E}\left\{\left[I(\Lambda_j(\boldsymbol{Z})<\varphi(c_{\alpha}))-
        I(T_j(\boldsymbol{Z})<c)\right]\left[\Pr(\theta_j=0\mid\boldsymbol{Z}, {\boldsymbol{\vartheta}})-\varphi(c_{\alpha})\Pr(\theta_j=1\mid\boldsymbol{Z}, {\boldsymbol{\vartheta}})\right]\right\}\leq0.\eqno{(\mathrm{A}.4)}
\]
The inequalities (A.3) and (A.4) yield that
\[
        (\varphi(\alpha)-\varphi(c_{\alpha}))\sum\limits^m_{j=1}\mathrm{E}\left\{[I(\Lambda_j(\boldsymbol{Z})<\varphi(c_{\alpha}))-I(T_j(\boldsymbol{Z})<c)]
        \Pr(\theta_j=1\mid\boldsymbol{Z}, {\boldsymbol{\vartheta}}) \right\}\leq0.\eqno{(\mathrm{A}.5)}
\]
Then it follows from
\[
        \sum\limits^m_{j=1}\mathrm{E}\left\{I(\Lambda_j(\boldsymbol{Z})<\varphi(c_{\alpha}))\left[\Pr(\theta_j=0\mid\boldsymbol{Z}, {\boldsymbol{\vartheta}})-\varphi(c_{\alpha})\Pr(\theta_j=1\mid\boldsymbol{Z}, {\boldsymbol{\vartheta}})\right]\right\}<0,
\]
and (A.1) that
\[
        \varphi(\alpha) = \frac{\sum\limits^m_{j=1}\mathrm{E}\left[I(\Lambda_j(\boldsymbol{Z})<\varphi(c_{\alpha}))\Pr(\theta_j=0\mid\boldsymbol{Z}, {\boldsymbol{\vartheta}})\right]}{\sum\limits^m_{j=1}\mathrm{E}\left[I(\Lambda_j(\boldsymbol{Z})<\varphi(c_{\alpha}))\Pr(\theta_j=1\mid\boldsymbol{Z}, {\boldsymbol{\vartheta}})\right]} < \varphi(c_{\alpha}).\eqno{(\mathrm{A}.6)}
\]
Then the inequalities (A.5) and (A.6) yield that
\[
         \sum\limits^m_{j=1}\mathrm{E}\left[I(\Lambda_j(\boldsymbol{Z})<\varphi(c_{\alpha}))\Pr(\theta_j=1\mid\boldsymbol{Z}, {\boldsymbol{\vartheta}}) \right] \geq \sum\limits^m_{j=1}\mathrm{E}\left[I(T_j(\boldsymbol{Z})<c)\Pr(\theta_j=1\mid\boldsymbol{Z}, {\boldsymbol{\vartheta}}) \right].
\]
Thus we have
\[
        \frac{1}{\sum\limits^m_{j=1}\mathrm{E}\left[\left(1-I(\Lambda_j(\boldsymbol{Z})<\varphi(c_{\alpha}))\right)\Pr(\theta_j=1\mid\boldsymbol{Z}, {\boldsymbol{\vartheta}}) \right]} \geq
        \frac{1}{\sum\limits^m_{j=1}\mathrm{E}\left[\left(1-I(T_j(\boldsymbol{Z})<c)\right)\Pr(\theta_j=1\mid\boldsymbol{Z}, {\boldsymbol{\vartheta}}) \right]}.\eqno{(\mathrm{A}.7)}
\]
It also follows from (A.4) that
\begin{eqnarray*}
  &&\sum\limits^m_{j=1}\mathrm{E}\left[\left(1-I(\Lambda_j(\boldsymbol{Z})<\varphi(c_{\alpha}))\right)\left(1-(1+\varphi(c_{\alpha}))
        \Pr(\theta_j=1\mid\boldsymbol{Z}, {\boldsymbol{\vartheta}})\right) \right]\\
  &\geq& \sum\limits^m_{j=1}\mathrm{E}\left[\left(1-I(T_j(\boldsymbol{Z})<c)\right)\left(1-(1+\varphi(c_{\alpha}))
        \Pr(\theta_j=1\mid\boldsymbol{Z}, {\boldsymbol{\vartheta}})\right) \right].
\end{eqnarray*}
Combining this with (A.7), we have that
\begin{eqnarray*}
    & & \frac{ \sum\limits^m_{j=1}\mathrm{E}\left[\left(1-I(\Lambda_j(\boldsymbol{Z})<\varphi(c_{\alpha}))\right)\left(1-(1+\varphi(c_{\alpha}))\Pr(\theta_j=1\mid\boldsymbol{Z}, {\boldsymbol{\vartheta}})\right) \right]}
    {\sum\limits^m_{j=1}\mathrm{E}\left[\left(1-I(\Lambda_j(\boldsymbol{Z})<\varphi(c_{\alpha}))\right)\Pr(\theta_j=1\mid\boldsymbol{Z}, {\boldsymbol{\vartheta}}) \right]} \\
    &\geq& \frac{\sum\limits^m_{j=1}\mathrm{E}\left[\left(1-I(T_j(\boldsymbol{Z})<c)\right)\left(1-(1+\varphi(c_{\alpha}))\Pr(\theta_j=1\mid\boldsymbol{Z}, {\boldsymbol{\vartheta}})\right) \right]}
    {\sum\limits^m_{j=1}\mathrm{E}\left[\left(1-I(T_j(\boldsymbol{Z})<c)\right)\Pr(\theta_j=1\mid\boldsymbol{Z}, {\boldsymbol{\vartheta}}) \right]}.
\end{eqnarray*}
Therefore,
\[
        \frac{1-(1+\varphi(c_{\alpha}))\mFNR\left({\bm\delta}(\Lambda(\boldsymbol{Z}),\varphi(c_{\alpha}))\right)}{\mFNR\left({\bm\delta}(\Lambda(\boldsymbol{Z}),\varphi(c_{\alpha}))\right)}
        \geq \frac{1-(1+\varphi(c_{\alpha}))\mFNR({\bm\delta}(T(\boldsymbol{Z}), c)}{\mFNR({\bm\delta}(T(\boldsymbol{Z}), c))}.
\]
Note that $\dfrac{1-(1+\varphi(c_{\alpha}))x}{x}$ is strictly decreasing in $x$ and ${\bm\delta}(\Lambda(\boldsymbol{Z}),\varphi(c_{\alpha}))={\bm\delta}(\HLIS(\boldsymbol{Z}),c_{\alpha})$, then we have
\[
        \mFNR\left({\bm\delta}(\HLIS(\boldsymbol{Z}),c_{\alpha})\right)\leq \mFNR({\bm\delta}(T(\boldsymbol{Z}), c)).
\]
\end{proof}

\par
\leftline{\bf Proof of Theorem 3}
\begin{proof}
The FDR of the oracle HLIS procedure (7) can be expressed
\begin{eqnarray*}
        \mathrm{FDR}_{HLIS} &=& \mathrm{E}_{\{\theta_i\}^m_{i=1}, \{Z_i\}^m_{i=1}}\left\{\dfrac{\sum\limits^m_{j=1}I(\HLIS_j(\boldsymbol{Z})<c^{*})(1-\theta_i)}{\sum\limits^m_{j=1}I(\HLIS_j(\boldsymbol{Z})<c^{*})}\right\}\\
                            &=& \mathrm{E}_{\{Z_i\}^m_{i=1}}\left\{\mathrm{E}\left[ \dfrac{\sum\limits^m_{j=1}I(\HLIS_j(\boldsymbol{Z})<c^{*})(1-\theta_i)}{\sum\limits^m_{j=1}I(\HLIS_j(\boldsymbol{Z})<c^{*})} \Bigg| \{Z_i\}^m_{i=1}\right]\right\}\\
                            &=& \mathrm{E}_{\{Z_i\}^m_{i=1}} \left\{\dfrac{\sum\limits^m_{j=1}I(\HLIS_j(\boldsymbol{Z})<c^{*})\HLIS_j(\boldsymbol{Z})}{\sum\limits^m_{j=1}I(\HLIS_j(\boldsymbol{Z})<c^{*})} \right\}
\end{eqnarray*}
Note that, for any $\{Z_i\}^m_{i=1}=\{z_i\}^m_{i=1}$, the cut-off $c^{*}$ satisfies the condition $\HLIS_{(l)}<c^{*}\leq\HLIS_{(l+1)}$, where
$$l=\max\left\{i:\frac{1}{i}\sum\limits^i_{j=1}\HLIS_{(j)}\leq\alpha\right\}.$$
Thus we have
\[
        \mathrm{FDR}_{HLIS} = \sum_{\{z_i\}^m_{i=1}}\left\{\frac{1}{l}\sum\limits^l_{j=1}\HLIS_{(j)}\right\}\Pr(\{Z_i\}^m_{i=1}=\{z_i\}^m_{i=1})\leq\alpha.
\]
\end{proof}

\par
For notational simplicity, let $\boldsymbol{\theta}=\{\theta_i\}^m_{i=1}$, $\boldsymbol{\eta}=\{\eta_i\}^m_{i=1}$, $\boldsymbol{z}=\{z_i\}^m_{i=1}$. Suppose that the underlying states and classes, $\{\theta_i\}^m_{i=1}$ and $\{\eta_i\}^m_{i=1}$, are observed, then the log-likelihood of the complete data can be expressed as:
{\small
\begin{eqnarray*}
\log L(\boldsymbol{\vartheta};\boldsymbol{\theta}, \boldsymbol{\eta}, \boldsymbol{z}) &=&  \sum_{k=1}^K I(\eta_1=k)\log \pi_k + \sum_{j=1}^{m-1} \sum_{k=1}^K \sum_{l=1}^K \left\{I(\eta_j=k, \eta_{j+1}=l)\log \left[\delta_{kl}^{s(j)} b_{kl}^{1-s(j)}\right]\right\}\\
   &&+  \sum_{j=1}^{m-1} \sum_{p=0}^1 \sum_{q=0}^1 \sum_{l=1}^K \left\{I(\theta_j=p, \theta_{j+1}=q, \eta_{j+1}=l)\log a_{pq}(l)\right\} \\
   &&+  \sum_{p=0}^1 \sum_{k=1}^K \left\{I(\theta_1=p, \eta_1=k)\log c_p(k)\right\} + \sum_{j=1}^{m} \sum_{p=0}^1 \left\{I(\theta_j=p) \log f_p(z_{j})\right\},
\end{eqnarray*}
}
where $I(\cdot)$ is an indicator function.

\par
Next, we will introduce in detail the EM algorithm \citep{baum1970a} used to calculate the maximum likelihood estimate. Denote by $\boldsymbol{\vartheta}^{(0)}=(\boldsymbol{\pi}^{(0)}, \boldsymbol{c}^{(0)}, \boldsymbol{\mathcal{A}}^{(0)}, \boldsymbol{\mathcal{B}}^{(0)}, \boldsymbol{\mathcal{F}}^{(0)})$ the initial parameters. At the E-step of the $t$-th iteration, the Q-function, defined as the expectation of the complete data log-likelihood given observations $\boldsymbol{z}$ and the current parameters $\boldsymbol{\vartheta}^{(t-1)}$, can be expressed as:
\begin{eqnarray*}
        Q(\boldsymbol{\vartheta}, \boldsymbol{\vartheta}^{(t-1)}) &=&  \E \left[\log L(\boldsymbol{\vartheta};\boldsymbol{\theta}, \boldsymbol{\eta}, \boldsymbol{z}) \mid \boldsymbol{z}, \boldsymbol{\vartheta}^{(t-1)} \right] \\
        &=& \sum_{k=1}^K \left\{\phi^{(t-1)}_1(k) \log \pi_k \right\} + \sum_{j=1}^{m-1} \sum_{k=1}^K \sum_{l=1}^K \left\{ \nu^{(t-1)}_j(k, l) \log \left[\delta_{kl}^{s(j)} b_{kl}^{1-s(j)}\right]\right\}\\
          &&+ \sum_{j=1}^{m-1} \sum_{p=0}^1 \sum_{q=0}^1 \sum_{l=1}^K \left\{\zeta^{(t-1)}_j(p, q, l) \log a_{pq}(l)\right\}\\
          &&+ \sum_{p=0}^1 \sum_{k=1}^K \left\{\rho^{(t-1)}_1(p, k) \log c_p(k)\right\} + \sum_{j=1}^{m} \sum_{p=0}^1 \left\{\gamma^{(t-1)}_j(p) \log f_p(z_{j})\right\},
\end{eqnarray*}
where $\phi^{(t-1)}_1(k)=\Pr(\eta_1=k \mid \{z_i\}^m_{i=1}, \boldsymbol{\vartheta}^{(t-1)})$, $\nu^{(t-1)}_j(k, l)=\Pr(\eta_j=k, \eta_{j+1}=l \mid \{z_i\}^m_{i=1}, \boldsymbol{\vartheta}^{(t-1)})$, $\zeta^{(t-1)}_j(p, q, k)=\Pr(\theta_j=p, \theta_{j+1}=q, \eta_{j+1}=k \mid \{z_i\}^m_{i=1}, \boldsymbol{\vartheta}^{(t-1)})$, $\rho^{(t-1)}_1(p, k)=\Pr(\theta_1=p, \eta_1=k \mid \{z_i\}^m_{i=1}, \boldsymbol{\vartheta}^{(t-1)})$ and $\gamma^{(t-1)}_j(p)=\Pr(\theta_j=p \mid \{z_i\}^m_{i=1}, \boldsymbol{\vartheta}^{(t-1)})$, for $p, q=0, 1$ and $k, l=1, \cdots, K$.
Denote by $\xi^{(t-1)}_j(p, q, k, l)$ the posterior probability of two consecutive states in the $t$-th iteration, that is, $\Pr(\theta_j=p, \theta_{j+1}=q, \eta_j=k, \eta_{j+1}=l \mid \{z_i\}^m_{i=1}, \boldsymbol{\vartheta}^{(t-1)})$. Some mathematical derivations yield that:
\[
        \xi^{(t-1)}_j(p, q, k, l) = \dfrac{\alpha^{(t-1)}_j(p, k)f^{(t-1)}_q(z_{j+1})\beta^{(t-1)}_{j+1}(q, l)a^{(t-1)}_{pq}(k) {\delta^{(t-1)}_{kl}}^{s(j)} {b^{(t-1)}_{kl}}^{1-s(j)}}{\sum\limits^1_{r=0}\sum\limits^1_{s=0}\sum\limits^K_{u=1}\sum\limits^K_{v=1}\left\{\alpha^{(t-1)}_j(r, u)f^{(t-1)}_s(z_{j+1})\beta^{(t-1)}_{j+1}(s, v)a^{(t-1)}_{rs}(u) {\delta^{(t-1)}_{uv}}^{s(j)} {b^{(t-1)}_{uv}}^{1-s(j)}\right\}}.
\]
Then the aforementioned variables can be expressed as:
{\small
\begin{eqnarray*}
        \phi^{(t-1)}_1(k)   &=& \sum\limits^1_{p=0}\sum\limits^1_{q=0}\sum\limits^K_{l=1}\xi^{(t-1)}_1(p, q, k, l), \\
        \nu^{(t-1)}_j(k, l) &=& \sum\limits^1_{p=0}\sum\limits^1_{q=0}\xi^{(t-1)}_j(p, q, k, l),\\
        \zeta^{(t-1)}_j(p, q, l) &=& \sum\limits^K_{k=1}\xi^{(t-1)}_j(p, q, k, l), \\
        \rho^{(t-1)}_1(p, k)     &=&  \sum\limits^1_{q=0}\sum\limits^K_{l=1}\xi^{(t-1)}_1(p, q, k, l), \\
        \gamma^{(t-1)}_j(p)      &=&  \sum\limits^1_{q=0}\sum\limits^K_{k=1}\sum\limits^K_{l=1}\xi^{(t-1)}_j(p, q, k, l),
\end{eqnarray*}
}
for $j=1, 2, \cdots, m-1$, $p, q=0, 1$ and $k, l=1, \cdots, K$.
\bibliographystyle{natbib}
\bibliography{Reference}

\begin{thebibliography}{}

\bibitem[{Baum} \emph{et~al.}(1970){Baum}, {Petrie}, {Soules}, and
  {Weiss}]{baum1970a}
{Baum}, L.~E., {Petrie}, T., {Soules}, G., and {Weiss}, N. (1970).
\newblock {A maximization technique occurring in the statistical analysis of
  probabilistic functions of Markov chains}.
\newblock \emph{Annals of Mathematical Statistics} \textbf{41}, 1, 164--171.

\bibitem[{Benjamini} and {Hochberg}(1995)]{benjamini1995controlling}
{Benjamini}, Y. and {Hochberg}, Y. (1995).
\newblock Controlling the false discovery rate: a practical and powerful
  approach to multiple testing.
\newblock \emph{Journal of The Royal Statistical Society Series B-statistical
  Methodology} \textbf{57}, 1, 289--300.

\bibitem[{Benjamini} and {Hochberg}(2000)]{benjamini2000on}
{Benjamini}, Y. and {Hochberg}, Y. (2000).
\newblock On the adaptive control of the false discovery rate in multiple
  testing with independent statistics.
\newblock \emph{Journal of Educational and Behavioral Statistics} \textbf{25},
  1, 60--83.

\bibitem[{Cui} \emph{et~al.}(2021){Cui}, {Wang}, and {Zhu}]{cui2021covariate}
{Cui}, T., {Wang}, P., and {Zhu}, W. (2021).
\newblock Covariate-adjusted multiple testing in genome-wide association
  studies via factorial hidden {Markov} models.
\newblock \emph{Test} \textbf{30}, 3, 737--757.

\bibitem[{Denti} \emph{et~al.}(2021){Denti}, {Guindani}, {Leisen}, {Lijoi},
  {Wadsworth}, and {Vannucci}]{Denti2021Two}
{Denti}, F., {Guindani}, M., {Leisen}, F., {Lijoi}, A., {Wadsworth}, W.~D., and
  {Vannucci}, M. (2021).
\newblock Two‐group poisson‐dirichlet mixtures for multiple testing.
\newblock \emph{Biometrics} \textbf{77}, 2, 622--633.

\bibitem[Efron(2007)]{efron2007correlation}
Efron, B. (2007).
\newblock Correlation and large-scale simultaneous significance testing.
\newblock \emph{Journal of the American Statistical Association} \textbf{102},
  477, 93--103.

\bibitem[{Efron} and {Tibshirani}(2002)]{efron2002empirical}
{Efron}, B. and {Tibshirani}, R. (2002).
\newblock Empirical bayes methods and false discovery rates for microarrays.
\newblock \emph{Genetic Epidemiology} \textbf{23}, 1, 70--86.

\bibitem[{Efron} \emph{et~al.}(2001){Efron}, {Tibshirani}, {Storey}, and
  {Tusher}]{efron2001empirical}
{Efron}, B., {Tibshirani}, R., {Storey}, J.~D., and {Tusher}, V. (2001).
\newblock Empirical {Bayes} analysis of a microarray experiment.
\newblock \emph{Journal of the American Statistical Association} \textbf{96},
  456, 1151--1160.

\bibitem[{Genovese} and {Wasserman}(2002)]{genovese2002operating}
{Genovese}, C. and {Wasserman}, L. (2002).
\newblock Operating characteristics and extensions of the false discovery rate
  procedure.
\newblock \emph{Journal of The Royal Statistical Society Series B-statistical
  Methodology} \textbf{64}, 3, 499--517.

\bibitem[{Genovese} and {Wasserman}(2004)]{genovese2004a}
{Genovese}, C. and {Wasserman}, L. (2004).
\newblock A stochastic process approach to false discovery control.
\newblock \emph{Annals of Statistics} \textbf{32}, 3, 1035--1061.

\bibitem[Hedenfalk \emph{et~al.}(2001)Hedenfalk, Duggan, Chen, Radmacher,
  Bittner, Simon, Meltzer, Gusterson, Esteller, and Raffeld]{2001Gene}
Hedenfalk, I., Duggan, D., Chen, Y., Radmacher, M., Bittner, M., Simon, R.,
  Meltzer, P., Gusterson, B., Esteller, M., and Raffeld, M. (2001).
\newblock Gene-expression profiles in hereditary breast cancer.
\newblock \emph{New England Journal of Medicine} \textbf{344}, 8, 539--548.

\bibitem[{Kuan} and {Chiang}(2012)]{kuan2012integrating}
{Kuan}, P.~F. and {Chiang}, D.~Y. (2012).
\newblock {Integrating prior knowledge in multiple testing under dependence
  with applications to detecting differential DNA methylation}.
\newblock \emph{Biometrics} \textbf{68}, 3, 774--783.

\bibitem[{Lichtenstein} \emph{et~al.}(2006){Lichtenstein}, {Björk}, {Hultman},
  {Scolnick}, {Sklar}, and {Sullivan}]{lichtenstein2006recurrence}
{Lichtenstein}, P., {Björk}, C., {Hultman}, C.~M., {Scolnick}, E., {Sklar},
  P., and {Sullivan}, P.~F. (2006).
\newblock Recurrence risks for schizophrenia in a swedish national cohort.
\newblock \emph{Psychological Medicine} \textbf{36}, 10, 1417--1425.

\bibitem[{Liu} \emph{et~al.}(2016){Liu}, {Zhang}, and {Page}]{liu2016multiple}
{Liu}, J., {Zhang}, C., and {Page}, D. (2016).
\newblock Multiple testing under dependence via graphical models.
\newblock \emph{The Annals of Applied Statistics} \textbf{10}, 3, 1699--1724.

\bibitem[{Marco} \emph{et~al.}(2017){Marco}, {Meuleman}, {Huang}, {Glass},
  {Pinello}, {Wang}, {Kellis}, and {Yuan}]{Marco2017multi}
{Marco}, E., {Meuleman}, W., {Huang}, J., {Glass}, K., {Pinello}, L., {Wang},
  J., {Kellis}, M., and {Yuan}, G.~C. (2017).
\newblock Multi-scale chromatin state annotation using a hierarchical hidden
  {Markov} model.
\newblock \emph{Nature Communications} \textbf{8}, 1, 1--9.

\bibitem[{Newton} \emph{et~al.}(2004){Newton}, {Noueiry}, {Sarkar}, and
  {Ahlquist}]{Newton2004Detecting}
{Newton}, M.~A., {Noueiry}, A., {Sarkar}, D., and {Ahlquist}, P. (2004).
\newblock Detecting differential gene expression with a semiparametric
  hierarchical mixture method.
\newblock \emph{Biostatistics} \textbf{5}, 2, 155--176.

\bibitem[{Owen}(2005)]{owen2005variance}
{Owen}, A.~B. (2005).
\newblock Variance of the number of false discoveries.
\newblock \emph{Journal of The Royal Statistical Society Series B-statistical
  Methodology} \textbf{67}, 3, 411--426.

\bibitem[{Ripke} \emph{et~al.}(2013){Ripke}, {O'Dushlaine}, and {Chambert K. et
  al.}]{ripke2013genome}
{Ripke}, S., {O'Dushlaine}, C., and {Chambert K. et al.} (2013).
\newblock Genome-wide association analysis identifies 14 new risk loci for
  schizophrenia.
\newblock \emph{Nature Genetics} \textbf{45}, 1150--1159.

\bibitem[{Schwartzman} and {Lin}(2011)]{schwartzman2011the}
{Schwartzman}, A. and {Lin}, X. (2011).
\newblock The effect of correlation in false discovery rate estimation.
\newblock \emph{Biometrika} \textbf{98}, 1, 199--214.

\bibitem[{Shu} \emph{et~al.}(2015){Shu}, {Nan}, and {Koeppe}]{shu2015multiple}
{Shu}, H., {Nan}, B., and {Koeppe}, R. (2015).
\newblock Multiple testing for neuroimaging via hidden markov random field.
\newblock \emph{Biometrics} \textbf{71}, 3, 741--750.

\bibitem[{Silverman}(2018)]{Silverman2018density}
{Silverman}, B.~W. (2018).
\newblock Density estimation for statistics and data analysis.
\newblock \emph{Routledge} .

\bibitem[Sun and Cai(2007)]{sun2007oracle}
Sun, W. and Cai, T.~T. (2007).
\newblock Oracle and adaptive compound decision rules for false discovery rate
  control.
\newblock \emph{Journal of the American Statistical Association} \textbf{102},
  479, 901--912.

\bibitem[{Sun} and {Cai}(2009)]{sun2009large}
{Sun}, W. and {Cai}, T.~T. (2009).
\newblock Large-scale multiple testing under dependence.
\newblock \emph{Journal of The Royal Statistical Society Series B-statistical
  Methodology} \textbf{71}, 2, 393--424.

\bibitem[{Sun} \emph{et~al.}(2015){Sun}, {Reich}, {Cai}, {Guindani}, and
  {Schwartzman}]{sun2015false}
{Sun}, W., {Reich}, B.~J., {Cai}, T.~T., {Guindani}, M., and {Schwartzman}, A.
  (2015).
\newblock False discovery control in large‐scale spatial multiple testing.
\newblock \emph{Journal of The Royal Statistical Society Series B-statistical
  Methodology} \textbf{77}, 1, 59--83.

\bibitem[{The International HapMap Consortium}(2003)]{HapMap3}
{The International HapMap Consortium} (2003).
\newblock The international hapmap project.
\newblock \emph{Nature} \textbf{426}, 789--796.

\bibitem[{Wang} and {Zhu}(2019)]{wang2019replicability}
{Wang}, P. and {Zhu}, W. (2019).
\newblock {Replicability analysis in genome-wide association studies via
  Cartesian hidden Markov models}.
\newblock \emph{BMC Bioinformatics} \textbf{20}, 1, 146.

\bibitem[{Wang} \emph{et~al.}(2019){Wang}, {Shojaie}, and
  {Zou}]{wang2019bayesian}
{Wang}, X., {Shojaie}, A., and {Zou}, J. (2019).
\newblock {Bayesian hidden Markov models for dependent large-scale multiple
  testing}.
\newblock \emph{Computational Statistics and Data Analysis} \textbf{136},
  123--136.

\bibitem[{Wei} \emph{et~al.}(2009){Wei}, {Sun}, {Wang}, and
  {Hakonarson}]{wei2009multiple}
{Wei}, Z., {Sun}, W., {Wang}, K., and {Hakonarson}, H. (2009).
\newblock Multiple testing in genome-wide association studies via hidden
  {Markov} models.
\newblock \emph{Bioinformatics} \textbf{25}, 21, 2802--2808.

\bibitem[Xiao \emph{et~al.}(2013)Xiao, Zhu, and Guo]{xiao2013large}
Xiao, J., Zhu, W., and Guo, J. (2013).
\newblock Large-scale multiple testing in genome-wide association studies via
  region-specific hidden markov models.
\newblock \emph{BMC Bioinformatics} \textbf{14}, 1, 282.

\bibitem[{Zhu} \emph{et~al.}(2012){Zhu}, {Jiang}, and
  {Zhang}]{zhu2012nonparametric}
{Zhu}, W., {Jiang}, Y., and {Zhang}, H. (2012).
\newblock Nonparametric covariate-adjusted association tests based on the
  generalized kendall's tau.
\newblock \emph{Journal of the American Statistical Association} \textbf{107},
  497, 1--11.

\end{thebibliography}


\begin{thebibliography}{}

\bibitem[Atkinson \emph{et~al.}(2007)Atkinson, Donev, and
  Tobias]{Atkinson2007Optimum}
Atkinson, A., Donev, A., and Tobias, R. (2007).
\newblock \emph{Optimum Experimental Designs, with SAS}.
\newblock Oxford University Press, New York.

\bibitem[Bertin-Mahieux \emph{et~al.}(2011)Bertin-Mahieux, Ellis, Whitman, and
  Lamere]{Bertin-Mahieux2011}
Bertin-Mahieux, T., Ellis, D.~P., Whitman, B., and Lamere, P. (2011).
\newblock The million song dataset.
\newblock In \emph{{Proceedings of the 12th International Conference on Music
  Information Retrieval ({ISMIR} 2011)}}.

\bibitem[Bertin-Mahieux \emph{et~al.}(2011)Bertin-Mahieux, Ellis, Whitman, and
  Lamere]{Bertin-Mahieux2011}
Bertin-Mahieux, T., Ellis, D.~P., Whitman, B., and Lamere, P. (2011).
\newblock The million song dataset.
\newblock In \emph{{Proceedings of the 12th International Conference on Music
  Information Retrieval ({ISMIR} 2011)}}.

\bibitem[Brown(1986)]{brown1986fundamentals}
Brown, L.~D. (1986).
\newblock \emph{Fundamentals of statistical exponential families: with
  applications in statistical decision theory}.
\newblock Lecture Notes-Monograph Series, vol. 9, Institute of Mathematical
  Statistics, Hayward, California.

\bibitem[Chapman \emph{et~al.}(1994)Chapman, Welch, Bowman, Sacks, and
  Walsh]{Chapman1994Arctic}
Chapman, W.~L., Welch, W.~J., Bowman, K.~P., Sacks, J., and Walsh, J.~E.
  (1994).
\newblock Arctic sea ice variability: Model sensitivities and a multidecadal
  simulation.
\newblock \emph{Journal of Geophysical Research Oceans} \textbf{99}, 919--935.

\bibitem[Chen(2011)]{Chen2011Quasi}
Chen, X. (2011).
\newblock \emph{Quasi Likelihood Method for Generalized Linear Model~(in
  Chinese)}.
\newblock Press of University of Science and Technology of China, HeFei.

\bibitem[Clemencon \emph{et~al.}(2014)Clemencon, Bertail, and
  Chautru]{clemencon2014}
lémencon, S., Bertail, P., \& Chautru, E. (2014).
\newblock Scaling up M-estimation via sampling designs: The Horvitz-Thompson stochastic gradient descent.
\newblock \emph{2014 IEEE International Conference on Big Data}  25-30.

\bibitem[Deb and Trivedi(1997)]{Deb1997DEMAND}
Deb, P. and Trivedi, P.~K. (1997).
\newblock Demand for medical care by the elderly: A finite mixture approach.
\newblock \emph{Journal of Applied Econometrics} \textbf{12}, 313--336.

\bibitem[Drineas \emph{et~al.}(2006{a})Drineas, Kannan, and
  Mahoney]{drineas2006fast}
Drineas, P., Kannan, R., and Mahoney, M.~W. (2006{a}).
\newblock Fast monte carlo algorithms for matrices i: Approximating matrix
  multiplication.
\newblock \emph{SIAM Journal on Computing} \textbf{36}, 132--157.

\bibitem[Drineas \emph{et~al.}(2011)Drineas, Magdon-Ismail, Mahoney, and
  Woodruff]{Drineas2011Fast}
Drineas, P., Magdon-Ismail, M., Mahoney, M.~W., and Woodruff, D.~P. (2011).
\newblock Fast approximation of matrix coherence and statistical leverage.
\newblock \emph{Journal of Machine Learning Research} \textbf{13}, 3475--3506.

\bibitem[Drineas \emph{et~al.}(2006{b})Drineas, Mahoney, and
  Muthukrishnan]{drineas2006sampling}
Drineas, P., Mahoney, M.~W., and Muthukrishnan, S. (2006{b}).
\newblock Sampling algorithms for $l_2$ regression and applications.
\newblock \emph{Proceedings of the Seventeenth Annual ACM-SIAM Symposium on
  Discrete Algorithm}  1127--1136.

\bibitem[{Drineas} \emph{et~al.}(2011){Drineas}, {Mahoney}, {Muthukrishnan},
  and {Sarl\'{o}s}]{Drineas2011Faster}
{Drineas}, P., {Mahoney}, M.~W., {Muthukrishnan}, S., and {Sarl\'{o}s}, T.
  (2011).
\newblock Faster least squares approximation.
\newblock \emph{Numerische Mathematik} \textbf{117}, 219--249.

\bibitem[Efron and Hastie(2016)]{Efron2016}
Efron, B. and Hastie, T. (2016).
\newblock \emph{Computer Age Statistical Inference: Algorithms, Evidence, and
  Data Science}.
\newblock Cambridge University Press, New York.

\bibitem[Fahrmeir and Kaufmann(1985)]{Fahrmeir1985}
Fahrmeir, L. and Kaufmann, H. (1985).
\newblock Consistency and asymptotic normality of the maximum likelihood estimator in generalized linear models.
\newblock \emph{The Annals of Statistics} \textbf{13}, 342-368.

\bibitem[Fahrmeir and Tutz(2001)]{fahrmeir2001multivariate}
Fahrmeir, L. and Tutz, G. (2001).
\newblock \emph{Multivariate statistical modelling based on generalized linear
  models}.
\newblock Springer, New York.

\bibitem[Ferguson(1996)]{Ferguson1996A}
Ferguson, T.~S. (1996).
\newblock \emph{A Course in Large Sample Theory}.
\newblock Chapman \& Hall, London.

\bibitem[Fonollosa \emph{et~al.}(2015)Fonollosa, Sheik, Huerta, and
  Marco]{Fonollosa2015Reservoir}
Fonollosa, J., Sheik, S., Huerta, R., and Marco, S. (2015).
\newblock Reservoir computing compensates slow response of chemosensor arrays
  exposed to fast varying gas concentrations in continuous monitoring.
\newblock \emph{Sensors and Actuators B Chemical} \textbf{215}, 618--629.

\bibitem[Goodson(2011)]{Goodson2011Mathematical}
Goodson, D.~Z. (2011).
\newblock \emph{Mathematical methods for physical and analytical chemistry}.
\newblock Wiley, Hoboken, New Jersey.

\bibitem[Hansen and Hurwitz(1943)]{Hansen1943On}
Hansen, M.~H. and Hurwitz, W.~N. (1943).
\newblock On the theory of sampling from finite populations.
\newblock \emph{The Annals of Mathematical Statistics} \textbf{14}, 2111--2118.

\bibitem[Jia \emph{et~al.}(2014)Jia, Michael, Petros, and
  Bin]{Jia2014Influence}
Jia, J., Michael, M., Petros, D., and Bin, Y. (2014).
\newblock Influence sampling for generalized linear models.
\newblock \emph{MMDS Foundation Technical report} .

\bibitem[Lee(1987)]{Lee1987Diagnostic}
Lee, A.~H. (1987).
\newblock Diagnostic displays for assessing leverage and influence in
  generalized linear models.
\newblock \emph{Australian Journal of Statistics} \textbf{29}, 233--243.

\bibitem[Loeppky \emph{et~al.}(2009)Loeppky, Sacks, and
  Welch]{Loeppky2009Choosing}
Loeppky, J.~L., Sacks, J., and Welch, W.~J. (2009).
\newblock Choosing the sample size of a computer experiment: A practical guide.
\newblock \emph{Technometrics} \textbf{51}, 366--376.

\bibitem[Ma \emph{et~al.}(2015)Ma, Mahoney, and Yu]{Ma2015A}
Ma, P., Mahoney, M.~W., and Yu, B. (2015).
\newblock A statistical perspective on algorithmic leveraging.
\newblock \emph{Journal of Machine Learning Research} \textbf{16}, 861--919.

\bibitem[Mahoney(2012)]{Mahoney2012Randomized}
Mahoney, M.~W. (2012).
\newblock Randomized algorithms for matrices and data.
\newblock \emph{Foundations and Trends in Machine Learning} \textbf{3},
  647--672.

\bibitem[Mccullagh and Nelder(1989)]{Mccullagh1989Generalized}
Mccullagh, P. and Nelder, J.~A. (1989).
\newblock \emph{Generalized Linear Models. Monographs on Statistics and Applied
  Probability 37}.
\newblock Chapman \& Hall, London.

\bibitem[Mersmann(2018)]{Olaf2018microbenchmark}
Mersmann, O. (2018).
\newblock \emph{microbenchmark: Accurate Timing Functions}.
\newblock R package version 1.4-4.

\bibitem[Pukelsheim(2006)]{pukelsheim2006optimal}
Pukelsheim, F. (2006).
\newblock \emph{Optimal design of experiments}.
\newblock Society for Industrial and Applied Mathematics, Philadelphia.

\bibitem[{R Core Team}(2018)]{Rpackage2018}
{R Core Team} (2018).
\newblock \emph{R: A Language and Environment for Statistical Computing}.
\newblock R Foundation for Statistical Computing, Vienna.

\bibitem[S\"{a}rndal \emph{et~al.}(1992)S\"{a}rndal, Swensson, and
  Wretman]{Sarndal1992Model}
S\"{a}rndal, C.~E., Swensson, B., and Wretman, J. (1992).
\newblock \emph{Model assisted survey sampling}.
\newblock Springer-Verlag, New York.

\bibitem[Silvapulle(1981)]{silvapulle1981existence}
Silvapulle, M.~J. (1981).
\newblock On the existence of maximum likelihood estimators for the binomial
  response models.
\newblock \emph{Journal of the Royal Statistical Society. Series B
  (Methodological)}  310--313.

\bibitem[van~der Vaart(1998)]{Vaart2000Asymptotic}
van~der Vaart, A. (1998).
\newblock \emph{Asymptotic statistics}.
\newblock Cambridge University Press, London.

\bibitem[Venables and Ripley(2002)]{Venables2002Modern}
Venables, W. N. and Ripley, B. D. (2002).
\newblock \emph{Modern Applied Statistics with S. Fourth Edition}.
\newblock Springer,  New York.


\bibitem[Wang \emph{et~al.}(2018)Wang, Yang, and Stufken]{Wang2017Information}
Wang, H.~Y., Yang, M., and Stufken, J. (2018).
\newblock Information-based optimal subdata selection for big data linear
  regression.
\newblock \emph{Journal of the American Statistical Association}
  doi:10.1080/01621459.2017.1408468.

\bibitem[Wang \emph{et~al.}(2018)Wang, Zhu, and Ma]{Wang2017Optimal}
Wang, H.~Y., Zhu, R., and Ma, P. (2018).
\newblock Optimal subsampling for large sample logistic regression.
\newblock \emph{Journal of the American Statistical Association}
\textbf{113}(522), 829--844.

\bibitem[Xiong and Li(2008)]{xiong2008some}
Xiong, S. and Li, G. (2008).
\newblock Some results on the convergence of conditional distributions.
\newblock \emph{Statistics and Probability Letters} \textbf{78}, 3249--3253.

\bibitem[Zhang and Jia(2017)]{zhang2017elastic}
Zhang, H. and Jia, J. (2017).
\newblock Elastic-net regularized high-dimensional negative binomial
  regression: Consistency and weak signals detection.
\newblock \emph{arXiv preprint arXiv:1712.03412} .

\bibitem[Zhu and Jiang(2018)]{zhu2018jiang}
Zhu, R. and Jiang, J. (2018).
\newblock Subsampled Optimization: Statistical Guarantees, Mean Squared Error Approximation, and Sampling Method.
\newblock \emph{arXiv preprint arXiv:1804.03615}.

\end{thebibliography}

\if1\blind
{

} \fi

\end{document}